\documentclass{elsart}

\usepackage{graphicx}
\usepackage{amsmath}
\usepackage{amssymb}
\usepackage{latexsym}

\newcommand{\graduw}{\nabla_U W}
\newcommand{\gradwu}{\nabla_W U}
\newcommand{\kg}{\kappa_g}
\newcommand{\ka}{\kappa}
\newcommand{\kd}{\kappa_d}

\newcommand{\hdu}{\hat{\delta u}}
\newcommand{\du}{\delta u}
\newcommand{\hcs}{\hat{c}}
\newcommand{\ccs}{\check{c}}
\newcommand{\etaL}{\eta_L}
\newcommand{\etar}{\eta_R}
\newcommand{\etas}{\eta_*}
\newcommand{\ttal}{\theta_L}
\newcommand{\ttar}{\theta_R}
\newcommand{\ttas}{\theta_*}

\begin{document}
\begin{frontmatter}

\title{A Hybrid Scheme for Gas-Dust Systems Stiffly Coupled via Viscous Drag}
\author{Francesco Miniati}
\address{Physics Department, Wolfgang-Pauli-Strasse 27, ETH Z\"urich, CH-8093 Z\"urich}

\begin{abstract}
  We present a stable and convergent method for studying a system of
  gas and dust, coupled through viscous drag in both non-stiff and
  stiff regimes. To account for the effects of dust drag in the update
  of the fluid quantities, we employ a fluid description of the dust
  component and study the modified gas-dust hyperbolic system
  following the approach in Miniati \& Colella (2007). In 
  addition to two entropy waves for the gas and dust components, 
  respectively, the extended system includes three waves driven
  partially by gas pressure and partially by dust drift, which,  in the 
  limit of vanishing coupling, tend to the two original acoustic waves 
  and the unhindered dust streaming. Based on this
  analysis we formulate a predictor step providing first order
  accurate reconstruction of the time-averaged state variables at cell
  interfaces, whence a second order accurate estimate of the
  conservative fluxes can be obtained through a suitable linearized
  Riemann solver.  The final source term update is carried out using a
  one-step, second order accurate, L-stable, predictor corrector
  asymptotic method (the $\alpha$-QSS method suggested by Mott
  et. al. 2000). This procedure completely defines a two-fluid method
  for gas-dust system. Using the updated fluid solution allows us to
  then advance the individual particle solutions, including
  self-consistently the time evolution of the gas velocity in the
  estimate of the drag force. This is done with a suitable particle
  scheme also based on the $\alpha$-QSS method. A set of benchmark
  problems shows that our method is stable and convergent. When dust
  is modeled as a fluid (two-fluid) second order accuracy is achieved
  in both stiff and non-stiff regimes, whereas when dust is modeled
  with particles (hybrid) second order is achieved in the non-stiff
  regime and first order otherwise.
\end{abstract}
\end{frontmatter}

\section{Introduction} \label{eq.sec}

We wish to solve the system of partial differential equations
describing two systems coupled through the exchange of momentum
through a viscous term proportional to their relative velocity.
This situation characterizes a variety of problems,
among others gas-dust coupling in protoplanetary disks, the motion of
polymer molecules in biological fluids~\cite{trebotich05}, 
drift of different ion species in the planetary plasma~\cite{birnetal01}.
We are interested in addressing the case in which the viscous coupling
becomes stiff, such that the relaxation time characterizing it is
significantly shorter than the smallest timescale characterizing the
fluid system, typically defined as sound crossing time of a resolution
element.  Although the results presented in this paper can possibly be
extended to other physical systems as those mentioned above, in the
following we specialized our analysis to the case of a gaseous and a
dust component, coupled through a drag term as well as gravity, typical
of a protoplanetary disk.
Without loss of generality we start considering the problem in one
dimension.  The gas component is described by the equations of
hydrodynamics with a suitable source term, namely
\begin{equation} \label{hypsys:eq}
\frac{\partial U}{\partial t} + \frac{\partial F(U)}{\partial x}  = S(U),
\end{equation}
where
\begin{equation} \label{fluid:eq}
U=
\begin{pmatrix}
\rho_g \\
\rho_g u_g \\
\rho_g E 
\end{pmatrix}, \quad
F(U)=
\begin{pmatrix}
\rho_g u_g \\
\rho_g u_g^2 + P \\
u_g\,[\rho E + P]
\end{pmatrix}, \quad
S(U)=
\begin{pmatrix}
0 \\
\rho_g(f_d -\nabla \phi)\\
u_g\rho_g(f_d - \nabla \phi) \\
\end{pmatrix}.
\end{equation}
The fluid quantities (with subscript $g$ for gas) have their usual
meaning, whereas $f_d$ and $-\nabla \phi$ describe drag and gravitational
acceleration, respectively, which will be specified below.  
The dust particles move along the following trajectories in phase-space
\begin{eqnarray} \label{dxdt:eq} 
\frac{d x_d}{dt} &=& v_d, \\ \label{dvdt:eq}
\frac{dv_d}{dt}&=& - \kd \,(v_{d}-u_{g}) -\nabla \phi,
\end{eqnarray}
where, $x_d$ and $v_d$ are the dust particles position and velocity,
respectively. We consider particles sizes that are small compared to
the gas particle mean free path (Epstein's regime) so that the drag
coefficient is
\begin{equation} \label{kd:eq}
\kd =\kappa_0  \rho_g c ,\quad \kappa_0 \equiv  \frac{1}{\hat\rho_ds}.
\end{equation}
Here, $\hat\rho_d$ and $s$ indicate the dust grain's mass density and
size, respectively, and $c$ is the gas speed of sound.  As shown below
(Sec.~\ref{tfde:se}), the back-reaction exerted by the dust particles
on the gas in Eq. (\ref{fluid:eq}), takes the form
\begin{equation} \label{kg0:eq}
f_d = -\kg(u_g-u_d),\quad \kg =\kappa_0  \rho_d c ,
\end{equation}
where $u_d$ is the average dust velocity in the neighborhood
of the considered fluid element, also to be defined below.
Eq. (\ref{kd:eq}) and (\ref{kg0:eq}) indicate that 
the dust-gas coupling coefficients become very large in 
the limit of either/both small grain sizes or/and large dust density.
In this case the relaxation between gas and dust is fast compared
to the sound crossing time, which leads to numerically stiff conditions. 
In addition, in the case of large dust densities, 
the dust back-reaction on the gas dynamics is considerable and needs
to be accounted for accurately.

In addition to the stiff conditions mentioned above, the problem is
complicated by the fact that on the one hand the dust particles
are collisionless and therefore best described with a particle method
in phase-space. However, in the stiff regime, the particles
effectively interact on a short timescale not only with the
surrounding gas but also with the surrounding particles. Taking into
account such interactions using directly a particle description
complicates considerably the problem, to a level that in fact is not
tractable.

In this paper, we formulate an algorithm to address such a problem.
In particular, for the purpose of modeling efficiently the effects of
the drag force by the dust particles on the fluid, we first obtain
fluid description of the dust particle component, using a
Particle-Mesh method.  We then define an extended conservative system
that includes both the gas and dust variables.  This systems is
advanced one time step following a method which is an extension of the
one described in Miniati and Colella~\cite{mico07a}.  More
specifically, we use the fluid description of the dust to account for
the modifications of the drag terms on the hyperbolic structure of the
gas equations. This allows us to formulate a predictor step that gives
first order accurate reconstruction of the time-averaged state
variables at cell interfaces, whence a second order accurate estimates
of the conservative fluxes can be obtained.  With the dust component
still described as a fluid, and a second order estimate of the fluxes
for our extended conservative system, the source update is finally
carried out using a one-step, second order accurate, predictor
corrector asymptotic method as suggested by Mott, Oren and van
Leer~\cite{Mottetal2000}.  At the end of this procedure we have
obtained a fluid solution that includes self-consistently the
collective effect of the particles drag on the fluid.

We can then update the individual particle solutions using a suitable
particle scheme that follows the particles along their characteristic
trajectories in phase space, taking into account the effect of drag
from the fluid. The above knowledge of the fluid solution at the
current and next time-step is crucial, however, to include
self-consistently the time evolution of the gas velocity in the
estimate of the drag force.

Finally, dust and gas are also coupled through gravity. However, this
can be achieved in the standard way (e.g.~\cite{mico07a}), namely by
applying to both dust and gas components the gradient of their
cumulative potential, $\phi$, defined by the following Poisson's
equation
\begin{equation}
\nabla^2 \phi = -4\pi {\mathcal G}(\rho_g+\rho_d).
\end{equation}
Several codes have already been published in the literature in order
to study the coupled dust-gas-dynamics in protoplanetary disks. Among
others, numerical codes based on the Smoothed Particle Hydrodynamics
method (SPH) approach have been developed for multifluid system
\cite{moko95,fouchet05,laibe08}, in addition to the simpler case in
which dust is treated as a test particle and backreaction is
neglected~\cite{riceetal04}.  Grid based two fluid methods have been
around for a long time~\cite{cuzzietal93}.  More recently they have been
extended to the MHD case~\cite{frone09}, and to one-fluid models in
which dust and gas are perfectly coupled~\cite{barranco09}.  Grid
based hybrid (fluid+particles) methods have also been developed,
including higher oder (sixth) spectral
methods~\cite{johansen06} and higher order (second) Godunov's
method~\cite{balsara09}.  The novelty of our method consists in its
ability to handle a variety of numerically stiff conditions both in
the two-fluid and hybrid approaches.
This is relevant because in a realistic setting stiff conditions arise
in limited regions of a system and at a certain point in time during
its evolution.
This paper is organized as follows. In Sec.~\ref{pis:se}.
the particle integration scheme is presented.
In Sec.~\ref{sdc:se} we describe
the two-fluid approach that allows us to update the fluid solution
one time step. This section, therefore includes a description of the
methods for the fluid treatment of the dust component, the
details of the semi-implicit predictor-corrector used for the source
update. 
In Sec.~\ref{tfps:se} we present the Godunov predictor step,
derive the characteristic analysis and provide a linearized Riemann
solver, for the extended gas-dust fluid system.  The stability of the
method is addressed in Sec.~\ref{sa:se}. Accuracy and convergence tests
are presented in Sec. \ref{tests:se} and a summary with discussion in Sec.
\ref{concl:se} concludes the paper.
\section{Particle Integration Scheme} \label{pis:se}

The particle positions and velocities is updated with a predictor
corrector method based on a variant of the Quasi-Steady-State method
proposed by Mott, Oran and van Leer~(the so called
$\alpha-$QSS,~\cite{Mottetal2000}).  Because we use this scheme
extensively in this paper, we first describe it briefly in the following.

Consider the first order ordinary differential equation (ODE)
\begin{equation} \label{sample:eq}
\frac{dy}{dt}=-p(t,y)y+q(t,y),
\end{equation}
\begin{equation}
y(t_0)=y_0,
\end{equation}
with $y\in\mathbb{R}$, 
$p,q:\mathbb{R}\times\mathbb{R}\rightarrow\mathbb{R}$. 
If $p$ and $q$ are constants, the above system has the following exact
solution given by Duhamel's formula,
\begin{equation} \label{duhaODE:eq}
y(t)=y_0 e^{-pt}+\frac{q}{p}(1-e^{-pt}).
\end{equation}
QSS methods are based on the asymptotic behavior of the above
solution.  When $p$ and $q$ depend on $(y,t),$ a first-order algorithm
can be obtained by setting $p=p_0\equiv p(t_0),~q=q_0\equiv q(t_0)$.
This approach corresponds to the simplest QSS method and can be used
to develop higher order methods by incorporating to some degree
the time dependence of $p,q$ into the solution.  All QSS methods,
however, reproduce the exact solution as $p,q$ become constant. There
are several QSS derived methods in the literature
(cf.~\cite{OranBoris2000} for a review).  Here we shall adopt the
$\alpha-$QSS method of Mott et al.~\cite{Mottetal2000}.  This is a
single step, second order accurate, A-stable predictor-corrector
method that can be summarized with the following procedure:
\begin{gather} \label{qsspred:eq}
\tilde{y}(t_0+\Delta t) = y_0 +\Delta t\frac{q_0-p_0y_0}{1+\alpha(p_0\Delta t) p_0\Delta t},\\ \label{qsscorr:eq}
y(t_0+\Delta t) = y_0 +\Delta t\frac{q_*-\bar p y_0}{1+\alpha( \bar p\Delta t) \bar p\Delta t},\\ \label{qssqparams:eq}
q_* = q_0+\alpha(\bar p\Delta t)~ [q(\tilde y,t_0+\Delta t)-q_0], \\ \label{qsspparams:eq}
\alpha(x) = \frac{1-\frac{1-e^{-x}}{x}}{1-e^{-x}},~\bar p=\frac{p_0+p(\tilde y,t_0+\Delta t)}{2}.
\end{gather}
Eq.~(\ref{qsspred:eq}) and (\ref{qsscorr:eq}) correspond to the
predictor and corrector step respectively.  In addition to the case
where $p,q$ are constants, the above algorithm also returns the exact
solution when $p$ is constant and $q$ is linear in time or $p$ is
linear in time and $q=0$~\cite{Mottetal2000}. 

In the following we
derive a predictor-corrector algorithm based on
Eq.~(\ref{qsspred:eq})-(\ref{qsscorr:eq}) to integrate
the equation of motion of the dust particles given by (\ref{dxdt:eq}) 
and (\ref{dvdt:eq}). 
In the absence of drag it effectively reduces to a kick-drift-kick variant of the leapfrog
scheme. For its derivation we assume that the gas velocity solution, $v_g$,
has already been computed by the fluid scheme described below, 
at both $t^n$ and $t^{n+1}=t^n+\Delta t$.  
The drag coefficient in (\ref{dvdt:eq}) is a function of the
gas density and sound speed and is interpolated at the particle position 
with a Particle-Mesh method, i.e.
\begin{equation} \label{kxp:eq}
\kd ({\bf x_p}) = \sum_{p} w[x_p - x({\bf i})] \kd({\bf i}),
\end{equation}
where $w(s)$ is a weight function, $x({\bf i})$, is the ${\bf i}$-th
cell's center, and the summation is carried out over all cells.
The velocity of the gas entering Eq.
(\ref{dvdt:eq}) must also be obtained by way of interpolation.
By using the following expression
\begin{equation} \label{ugxp:eq}
v_g ({\bf x_p})  = \frac{1}{\kd ({\bf x_p})} \sum_{p} w[x_p - x({\bf i})] \kd({\bf i}) u_g({\bf i}),
\end{equation}
total dust+gas momentum is conserved by construction in the non-stiff limit,
and simple tests show that it is also conserved with high accuracy in the stiff 
regime.

We now present the algorithm, but a more detailed description of its derivation
is provided in Appendix \ref{paqss:app}.
We first predict the particle position at $t^{n+1}$ as
\begin{equation}\label{xtlde:eq}
\tilde x_d = x_d^n + \beta(\kd^n\Delta t)v_d^n\Delta t
+[1-\beta(\kd^n\Delta t)]v_g^n\Delta t +\frac{1}{2}\nabla\phi^n\Delta t^2,
\end{equation}
where $\beta(x) = \frac{1-e^{-x}}{x}$.
As the particle travels from $x_d^n$ to $\tilde x_d$ the gas velocity
changes from $v_g^n=v_g(x_d^n)$ to $v_g^{n+1}=v_g(\tilde x_d)$.
Since the above change is partially due to drag relaxation and
partially due to the motion of the particles across a velocity gradient,
we assume that the velocity experienced by the particle evolves as
follows:
\begin{equation} \label{vgoft:eq}
v_g(t)= v_g^{n} + \Delta v_g 
\left(e^{-\kg\Delta t} \frac{t}{\Delta t}+ 1-e^{-\kg t}\right),
\end{equation}
where, $\Delta v_g\equiv v_g^{n+1}- v_g^{n}$. 
The above expression allows for a relaxation of the gas velocity
towards the value $v_g^{n+1}$ much faster then linear, which is
important in the stiff case ($\kg\Delta t\gg 1$).  For the stiff case,
this introduces some differences in our integration scheme with
respect to the pure $\alpha$-QSS method.

With the assumption of Eq.~(\ref{vgoft:eq}) we can carry out the time
integration of the equation of motions for the particle velocity and
position. In doing so, as in the $\alpha$-QSS approach, 
we replace the time dependent drag coefficients
with their time averages, that is the average between the values
at position $x_d^n$ and $\tilde x_d$, $\kappa_s\rightarrow\bar{\kappa_s}=
[\kappa_s(x_d^n)+\kappa_s(\tilde x_d)]/2$, $s=d,g$.
We thus obtain
\begin{gather} \nonumber 
x_d^{n+1} =x_d^n + v_d^{n} \Delta t + 
\left[ 1-\beta\left(\bar{\kd}\Delta t\right) \right]  
\left(v_g^n-v^n_d\right) \Delta t +\nabla\phi^n\frac{\Delta t^2}{2}
\\ + \Delta v_g \left\{e^{-\bar{\kg}\Delta t}
\left[\frac{1}{2}-\frac{1-\beta\left(\bar{\kd}\Delta t\right)}{\bar{\kd}\Delta t}
\right]+ 
1-\frac{\bar{\kd}\beta(\bar{\kg}\Delta t)-\bar{\kg}\beta(\bar{\kd}\Delta t)}{
\bar{\kd}-\bar{\kg}} \right\} \Delta t, \label{DxDt:eq}\\ 
v_d^{n+1} =v_d^{n} + \beta(\bar{\kd}\Delta t)\bar{\kd}(v_g^{n}-v_d^{n})\Delta t 
\nonumber\\
+\Delta v_g \left\{e^{-\bar{\kg}\Delta t}[1-\beta(\bar{\kd})]+
\frac{\bar{\kd}\bar{\kg}}{\bar{\kd}-\bar{\kg}}[\beta(\bar{\kg}\Delta t)-\beta(\bar{\kd}\Delta t)] \Delta t
\right\}.\label{DvDt:eq}
\end{gather}

Remarkably the above particle method contains no explicit term arising
from the stiff coupling of the particle component.  Of course, the
$\beta$ terms take into account that the gas-dust coupling is fast
compared to the timestep $\Delta t$. However, each particle motion is
integrated individually though it effectively depends on the other
particles solutions, and the scheme is essentially explicit in time,
as it only involves the particle solution at time $t=n\Delta t$.
Still the scheme is stable and convergent, and this is due to the fact
that the gas velocity solution at times $t$ and $t+\Delta t$, with
which the particles interact, already contains the effect of the dust
component to second order accuracy, even in the stiff regime.  The gas
velocity solution is in fact the only quantity entering
Eq.~(\ref{DxDt:eq})-(\ref{DvDt:eq}). We have actually tried to employ
more sophisticated approaches than Eq.~(\ref{vgoft:eq}), which would
also involve the dust fluid velocity, but despite the higher degree of
complexity and computational cost, they did not improve on the
algorithm accuracy.
\section{Two-Fluid Semi-Implicit Predictor-Corrector} \label{sdc:se}
\subsection{Two-Fluid Description} \label{tfde:se}

In order to efficiently model the collective effect of the drag force
exerted by the dust particles on the fluid we use a fluid like
description of the dust component.  Thus, using a Particle-Mesh method
we define the dust density and velocity field on a Cartesian grid as
\begin{eqnarray} \label{rhod:eq}
\rho_d ({\bf i}) & = & \frac{1}{\Delta x^3}\sum_{p} w[x_p - x({\bf i})] m_p, 
\\ \label{veld:eq}
u_d ({\bf i}) & = &\sum_{p} w[x_p - x({\bf i})] v_p,
\end{eqnarray}
where as in the previous section $w(s)$ is a weight function, $x({\bf i})$, is the ${\bf i}$-th
cell's center, but now the summation is carried out over all the dust
particles.  Note that the position and velocity of the dust particles
are updated independently at each time-step using a particle-method
described in Sec.~\ref{pis:se}. The time evolution of $\rho_d$ and
$\rho_du_d$ is then obtained from the zero-th and first velocity
moments, respectively, of Boltzmann's equation for the dust
distribution function, $g_d(x,v,t)$. Using the particle equation of
motion (\ref{dxdt:eq})-(\ref{dvdt:eq}), and the appropriate collision
term we obtain
\begin{eqnarray}\label{drddt:eq}
\frac{\partial\rho_d}{\partial t} + \frac{\partial}{\partial x}(\rho_d u_d)
 &=&0, \\ \label{dvddt:eq}
\frac{\partial}{\partial t}( \rho_d u_d)+ \frac{\partial}{\partial x} ( \rho_d u_d u_d)
 &=&
 - \rho_d \kd (u_d-u_g) - \rho_d\nabla \phi,
\end{eqnarray}
where closure is granted by neglecting higher order velocity terms as
suggested by~\cite{garaudetal2004}. However, different closures can in
principle be employed starting from the particle description as need be.
Finally, by Newton's third Law, Eq.~(\ref{dvddt:eq}) indicates that
the collective drag force exerted on the gas by the dust particles
ought to be
\begin{equation} \label{kg:eq}
f_d = -\kg(u_g-u_d),~~~\kg = \kd \frac{\rho_d}{\rho_g},
\end{equation}
ensuring momentum conservation.

We can now extend the set of conservative variables, $U$, to include
the density and momentum of the dust component,
\begin{equation}
U=(\rho,\rho u_g,\rho E)^T \rightarrow (\rho_g,\rho_g u_g,\rho E,\rho_d,\rho_d u_d)^T.
\end{equation}
The extended set of equations for $U$ is obtained by combining
(\ref{hypsys:eq})-(\ref{fluid:eq}) and (\ref{drddt:eq})-(\ref{dvddt:eq}).
After rearranging the source terms, it reads
\begin{equation} \label{exthypsys:eq}
\frac{\partial U}{\partial t} + \frac{\partial F(U)}{\partial x}  = K_U U + S(U),
\end{equation}
where
\begin{equation} \label{extfluid:eq}
F(U)=
\begin{pmatrix}
\rho u_g \\
\rho u_g^2 + P \\
u_g\,[\rho E + P] \\
\rho u_d \\
\rho u_d^2
\end{pmatrix}, \quad
S(U)=
\begin{pmatrix}
0 \\
\rho_g\nabla \phi\\
u_g \rho_g\nabla \phi \\
0 \\
\rho_d \nabla \phi
\end{pmatrix}, \quad
K_U=
\begin{pmatrix}
0 & 0 & 0  & 0 & 0  \\
0 & -\kg & 0 & 0 & \kd \\
0 & 0 & 0 & 0  & 0 \\
0 & 0 & 0 &  0 & 0 \\
0 & \kg & 0 & 0 & -\kd 
\end{pmatrix} .
\end{equation}

\subsection{Semi-Implicit Method} \label{sim:se}

Given our extended system of equations~(\ref{exthypsys:eq})
we aim for a scheme in which an explicit approach is retained for the
non-stiff conservative hydrodynamic term, $\nabla\cdot F$, and a
semi-implicit method is employed for the stiff part of the source terms.
As in Miniati \& Colella \cite{mico07a}, our time discretization for
the source terms is a single-step, second-order accurate scheme.
However, instead of a method based on the deferred correction
approach~\cite{dugrro00}, here we derive our predictor corrector 
using again the  $\alpha-$QSS~\cite{Mottetal2000}.  The
reason is that although stable and convergent, unlike the $\alpha-$QSS
method, the deferred correction method is not L-stable (although a
combination of two such methods can lead to L-stability).  This can
lead to large errors, particularly with regard to momentum conservation, 
in a hybrid method in which fluid and particles are stiffly coupled. 

Our semi-implicit method consists in solving the following collection
of ODEs, one at each grid point,
\begin{equation} \label{odes:eq}
\frac{d U}{d t} = K_UU+S(U) - (\nabla \cdot \vec{F})^{n + \frac{1}{2}},
\end{equation}
where we view the time-centered flux divergence as a constant source,
whose computation using a modified Godunov method is described below.
Our predictor-corrector step then reads:
\begin{gather} \label{pred:eq}
\tilde{U} (t_0+\Delta t) = e^{K_0\Delta t} 
U_0 + {\mathcal I}^0_{K_0}(\Delta t) 
\left[S(U_0) - (\nabla \cdot \vec{F})^{n + \frac{1}{2}}\right] 
\Delta t, \\ \nonumber
U(t_0+\Delta t) = e^{\bar K\Delta t} U_0+ 
{\mathcal I}^0_{\bar K}(\Delta t) 
\left[S(U_0) - (\nabla \cdot \vec{F})^{n + \frac{1}{2}}\right]  
\Delta t  \\ \label{corr:eq}
+ {\mathcal I}^1_{\bar K}(\Delta t) 
\frac{S(\tilde{U})-S(U_0)}{\Delta t}
\frac{\Delta t^2}{2},
\end{gather}
where, $\bar K\equiv [K(U_0)+K(\tilde U)]/2$,
is used in Eq.~(\ref{corr:eq}), 
and we have defined the set of operators
\begin{gather} \label{iobt:eq}
{\mathcal I}^n_O(t) \equiv \frac{n!}{t^n} \, \int_0^t e^{(t-\tau)O} \tau^n d\tau.
\end{gather}

Note that gravity is actually unaffected by the operators ${\mathcal
  I}^n_O(t)$; however the form of Eq.~(\ref{pred:eq})-(\ref{corr:eq})
is suitable for a more general form of $S(U)$, which may even include
stiff terms associated, for example, with
an endothermic source.\footnote{In this case, if $\Sigma$ indicates
  the endothermic source, Eq.~(\ref{odes:eq}), as well as
  Eq.~(\ref{pred:eq})-(\ref{corr:eq}), should be modified by
  replacing
\begin{gather}
S(U) \leftarrow S(U)+\Sigma(U_0), \\
K(U)\leftarrow K(U)+\nabla_U\Sigma(U).
\end{gather}
}
Although the form of predictor corrector in the above equations
appears different from the original system
(\ref{qsspred:eq})-(\ref{qsscorr:eq}), 
their equivalence can be easily verified by replacing the operators $K,Q$
with the coefficients $p,q$. It is also easy to see that in
the non-stiff limit, the above scheme reduces to the usual second
order accurate explicit formulation
\begin{equation}
U(t_0+\Delta t) =  (1+\bar K \Delta t) U_0 - \Delta t \, (\nabla\cdot F)^{n+\frac{1}{2}} +
\frac{\Delta t}{2} \left[ S(\tilde{U}) + S(U_0) \right].
\label{eq:nnstifflim}
\end{equation}

\section{Two-Fluid Predictor Step} \label{tfps:se}
In order to compute our predictor step,
we cast the extended two-fluid gas-dust system in primitive form as follows:
\begin{equation} \label{hypsys:prim:eq}
\frac{\partial W}{\partial t} +A(W)\frac{\partial W}{\partial x} = K \, W + G,
\end{equation}
where the primitive variables are
\begin{equation} \label{gasdst.eq}
W=\left(
\rho_g ,
u_g ,
P ,
\rho_d ,
u_d \right)^T,
\end{equation}
and the quasi-linear operator, $A(W)\equiv \graduw\cdot\nabla_U
F\cdot\gradwu$, which includes an advection and a Lagrangian
component, is
\begin{gather}
A(W) \equiv u_g {\rm I}+A_L , ~~~
A_L=\begin{pmatrix}
0 & \rho_g & 0  & 0 & 0  \\
0 & 0 & \rho_g^{-1} & 0 & 0 \\
0 & \rho_g c^2 & 0 & 0 & 0 \\
0 & 0 & 0 & (u_d-u_g) & \rho_d \\
0 & 0 & 0 & 0  & (u_d-u_g) 
\end{pmatrix} .
\end{gather}
Finally
\begin{equation} \label{keq:eq}
K=
\begin{pmatrix}
0 & 0 & 0  & 0 & 0  \\
0 & -\kg & 0 & 0 & \kg \\
0 & 0 & 0 & 0  & 0 \\
0 & 0 & 0 &  0 & 0 \\
0 & \kd & 0 & 0 & -\kd 
\end{pmatrix} ,
\quad 
G =
\begin{pmatrix}
0 \\
-\nabla \phi \\
0 \\
0 \\
-\nabla \phi\\
\end{pmatrix} ,
\end{equation}
represent the drag operator and the residual source term, including gravity,
in primitive form.
Following the method in Miniati \& Colella~\cite{mico07a} we can follow the
dynamics of the system along Lagrangian trajectories
\begin{equation}
\frac{DW}{Dt} = -A_L(W)\frac{\partial W}{\partial x} + K \, W + G,
\end{equation}
with solution given by Duhamel's formula:
\begin{equation}
W(t) = e^{Kt} W(0) -
\int_0^t e^{K(t-\tau)} \left[ A_L \frac{\partial W}{\partial x}- G\right]d\tau.
\end{equation}
This allows us to define a modified dynamics reading
\begin{equation}\label{dsol:eq}
\frac{DW}{D t} +
{\mathcal I}_{K}(t)  A_L \frac{\partial W}{\partial x} = 
{\mathcal I}_{K}(t) \left[K W(0)+ G\right] + O(t).
\end{equation}
\subsection{Characteristic Analysis} \label{charanalys:se}

We use the quasi-linear system~(\ref{dsol:eq}) with $ t= \Delta
t/2$ to compute the Godunov predictor step.  Thus we
analyze the modified hyperbolic structure of that system of equations.
With the choice of $K$ in~(\ref{keq:eq}), from Eq.~(\ref{iobt:eq}) we
obtain
\begin{equation}
e^{K\Delta t/2} = 
\begin{pmatrix}
1 & 0 & 0 & 0 & 0\\
0 & \frac{\kd+\kg e^{-\ka\Delta t/2}}{\ka} & 0 & 0 & \frac{\kg (1-e^{-\ka \Delta t/2})}{\ka} \\
0 & 0 & 1 & 0 & 0 \\
0 & 0 & 0 & 1 & 0 \\
0 &  \frac{\kd(1-e^{-\ka \Delta t/2})}{\ka} & 0 & 0 & \frac{\kg+\kd e^{-\ka \Delta t/2}}{\ka}
\end{pmatrix},
\end{equation}
where $\ka= \kg+\kd$ is the total relaxation rate, and
\begin{equation}
{\mathcal I}_{K}(\Delta t/2)=
\begin{pmatrix}
1 & 0 & 0 & 0 & 0\\
0 & \frac{\kd+\beta\kg}{\ka} & 0 & 0 &  \frac{\kg (1-\beta)}{\ka} \\
0 & 0 & 1 & 0 & 0 \\
0 & 0 & 0 & 1 & 0 \\
0 & \frac{\kd(1-\beta)}{\ka} & 0 & 0 & \frac{\kg+\beta \kd}{\ka}
\end{pmatrix},
\end{equation}
where
\begin{equation} \label{alpha:eq}
\beta=\beta\left(\frac{\ka\Delta t}{2}\right)=  \frac{1-e^{-\frac{1}{2}\ka\Delta t }}{\frac{1}{2}\ka\Delta t}, ~~0<\beta<1.
\end{equation}
We can then define the modified linear operator
\begin{equation}\label{aeff:eq}
{\mathcal I}_{K}(\Delta t/2) A_L = A^{\rm eff}_L =
\begin{pmatrix}
0 & \rho_g & 0 & 0 & 0 \\
0 & 0 & \frac{\kd+\beta\kg}{\ka\rho_g} & 0 & \frac{\kg(1-\beta)}{\ka} (u_d-u_g)\\
0 & \rho_g c^2 & 0 & 0  & 0 \\
0 & 0 & 0 & (u_d-u_g) & \rho_d \\
0 & 0 & \frac{\kd(1-\beta)}{\ka\rho_g} & 0 & \frac{\kg+\beta\kd}{\ka}(u_d-u_g)
\end{pmatrix},
\end{equation}
with associated characteristic equation
\begin{equation} \label{char::eq}
\lambda \left[\lambda-(u_d-u_g) \right]
\left[\lambda^3 -\lambda^2 \frac{\kg+\beta\kd}{\ka } (u_d-u_g)
-\lambda \frac{\kd+\beta\kg}{\ka} c^2+\beta c^2 (u_d-u_g)\right] = 0.
\end{equation}
The system eigenvalues are given by the roots of the above equation.
The first obvious solution, $\lambda^0=0$, correspond to an entropy
wave for the gas component. Since an entropy wave does not involve
velocity perturbations, we expect this type of wave to be unaffected
by the presence of a drag term.  The other obvious eigenvalue,
$\lambda^d=(u_d-u_g)$, is the equivalent of an entropy wave, but for
the dust component.  Perturbations in $\rho_d$ propagate only along
the characteristic curve associated to this eigenvalue and, therefore,
do not affect other primitive variables.
For this reason, in the following 
we simplify the characteristic analysis by dropping
out the $\rho_d$ components and assume that: $\rho_d$ is transported as
a passive scalar with speed  $\lambda^d$ and its intermediate state Riemann 
solution is simply given by the average of left and right state values.

The remaining eigenvalues of the system are given by the roots of the
cubic polynomial appearing in Eq.~(\ref{char::eq}), which can be shown
to be always distinct and real, with a few exceptions discussed below.
The set of relevant eigenvalues is then
\begin{eqnarray} 
\label{lambda0:eq}
\lambda^0 & = & 0, \\
\label{lambda1:eq}
\lambda^+&=&\frac{1}{3}\hat{\delta u}+ \frac{2}{\sqrt{3}} \left(\hcs^2 + \frac{\hdu^2}{3}\right)^{1/2}\cos\left(\frac{\varphi}{3}\right),\\
\lambda^-&=&\frac{1}{3}\hat{\delta u}+\frac{2}{\sqrt{3}} \left(\hcs^2 + \frac{\hdu^2}{3}\right)^{1/2}\cos\left(\frac{\varphi}{3}+ \frac{2}{3}\pi\right),
\label{lambda2:eq} \\
\lambda^\times &=&\frac{1}{3}\hat{\delta u}+\frac{2}{\sqrt{3}} \left(\hcs^2 + \frac{\hdu^2}{3}\right)^{1/2}\cos\left(\frac{\varphi}{3}+ \frac{4}{3}\pi\right),
\label{lambda3:eq}
\end{eqnarray}
where
\begin{eqnarray} \label{cduphi:eq}
\hcs&=&\sqrt{\frac{\kd + \beta\kg}{\ka}}\,c,~~\hat{\delta u} 
=\frac{\kg+\beta \kd}{\ka}\du,~~\du=u_d-u_g,\\
\varphi&=&\cos^{-1}\left[\hdu
\frac{2\hat{\delta u}^2+
9\hcs^2\left(1-\frac{3\beta\ka^2}{(\kd+\beta\kg)(\kg+\beta \kd)}\right)}{
2 (3\hcs^2 + \hdu^2 )^\frac{3}{2}}\right],\label{phi:eq}
\end{eqnarray}
\begin{figure*}
\begin{center}
\includegraphics[width=0.75\textwidth,angle=0]{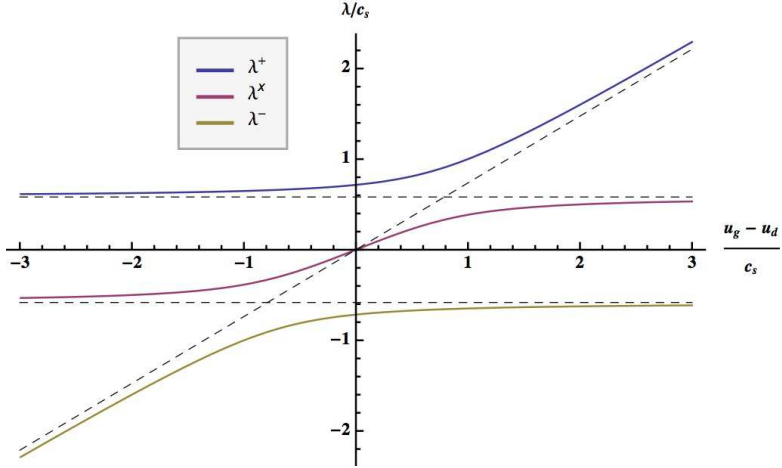}
\caption{
Solid lines: eigenvalues versus $(u_g-u_d)$, in sound speed units,
for $\beta=0.25$ and $\kg /\kappa = 0.65$. The dash lines indicate
the various asymptotic branches corresponding to $\lambda=\ccs$ (high, horizontal)
$\lambda=-\ccs$ (low horizontal) and $\lambda=\hdu$ (diagonal).
\label{ev:fig}}
\end{center}
\end{figure*}
and $\cos^{-1}$ conventionally indicates the principal value of the
multivalued inverse cosine function.  With this convention we always
have $\lambda^+>(\lambda^\times,\lambda^0)>\lambda^-$, whereas all
three cases $\lambda^0 \gtreqless \lambda^\times$ are
admitted.  An extra eigen-mode driven by the dust drift has appeared.
Under stiff dust-gas coupling conditions ($\beta<1$), this mode mixes
with the pressure driven modes thus acquiring an acoustic-like
character.  Likewise, the properties of the original sound waves are
contaminated by the dust component. The degree of mutual
contamination is regulated by the `mixing angle' $\varphi$.
In general sound speed is affected by both the additional inertia and
by the motion of the dust particles.
To see this in more details in the following
we explore the asymptotic behavior of the waves in a few cases
of interest. The general results are summarized in Fig.~\ref{ev:fig}.
In the non stiff limit, we have
\begin{equation} \label{nonstifflim:eq}
\lim_{\beta\to 1}\lambda^+= c,~~
\lim_{\beta\to 1}\lambda^-= -c,~~
\lim_{\beta\to 1}\lambda^\times= \du,
\end{equation}
thus recovering the expected values. However,
as the dust particle drift vanishes we have
\begin{equation}
\lim_{\du\to 0}\lambda^+= \hcs,~~
\lim_{\du\to 0}\lambda^-= -\hcs,~~
\lim_{\du\to 0}\lambda^\times= \frac{\beta\ka}{\kd + \beta\kg}\hdu  .
\end{equation}
Here $\hcs$ is given in Eq.(\ref{cduphi:eq}) and correspond to a 
sound speed in which the gas density is replaced by an effective
average between the dust and the gas densities. So, in the stiff limit
($\beta \rightarrow 0$), it is the total density that enters the
definition of the sound speed (which can be dominated by the dust 
density). Similarly, in the limit of large values of $\hdu$ we obtain
(see dashed lines in Fig.~\ref{ev:fig})
\begin{gather}
\lim_{\hdu\to \infty}\lambda^+= \hdu,~~
\lim_{\hdu\to \infty}\lambda^-=-\ccs,~~
\lim_{\hdu\to \infty}\lambda^\times= \ccs, \\
\lim_{\hdu\to -\infty}\lambda^+= \ccs,~~
\lim_{\hdu\to -\infty}\lambda^-=  \hdu,~~
\lim_{\hdu\to -\infty}\lambda^\times=-\ccs,
\end{gather}
where we have used
\begin{equation}
\ccs= \sqrt{\frac{\beta\ka}{\kg + \beta\kd}}c,
\end{equation}
which represent an effective sound speed when the perturbation
propagates opposite to the direction of the dust particles velocity.
The said propagation speed vanishes in the stiff limit, but it always
coincide with the sound speed ($c$) in the limit of negligible dust
density (i.e. $\kg\rightarrow 0,~\kd\rightarrow \ka$).

To conclude the analysis we introduce the array of right eigenvectors
(by column) 

\begin{equation} \label{rightev:eq}
 R \equiv \left(
 \begin{array}{cccc}
  1 & 1 & 1 & 1 \\
  \frac{\lambda^-}{\rho _g} & 0 & \frac{\lambda^\times}{\rho _g} & 
 \frac{\lambda^+}{\rho _g} \\
  c^2 & 0 & c^2 & c^2 \\
  \frac{(\hcs-\lambda^-)(\hcs+\lambda^-)}{\mu\du(\beta-1)} & 0 & 
  \frac{(\hcs-\lambda^\times)(\hcs+\lambda^\times)}{\mu\du (\beta-1)} & 
  \frac{(\hcs-\lambda^+)(\hcs+\lambda^+)}{\mu\du (\beta-1)}
 \end{array}
 \right),
\end{equation}
where, to simplify the notation, we have introduced the reduced density,
$\mu\equiv \rho_g\rho_d/(\rho_g+\rho_d)$. Similarly, the array of
left eigenvectors (by row) is
\begin{equation} \label{leftev:eq}
L\equiv\left(
\begin{array}{cccc}
 0 &
-\frac{\left(\lambda^++\lambda^\times\right) \rho _g}{\left(\lambda^+-\lambda^-\right)\left(\lambda^\times-\lambda^-\right)} & 
\frac{\hcs^2 +\lambda^+ \lambda^\times}{c^2\left(\lambda^+-\lambda^-\right)\left(\lambda^\times-\lambda^-\right)} &
 -\frac{\mu\du (\beta-1)}{\left(\lambda^+-\lambda^-\right)\left(\lambda^\times-\lambda^-\right)} \\
 1 & 0 & -\frac{1}{c^2} & 0 \\
 0 & \frac{\left(\lambda^++\lambda^-\right) \rho _g}{(\lambda^+-\lambda^\times)(\lambda^\times-\lambda^-)} &
-\frac{\hcs^2 +\lambda^+ \lambda^-}{c^2(\lambda^+-\lambda^\times)(\lambda^\times-\lambda^-)} & \frac{\mu\du (\beta-1)}{(\lambda^+-\lambda^\times)(\lambda^\times-\lambda^-)} \\
 0 & -\frac{\left(\lambda^\times+\lambda^-\right) \rho _g}{\left(\lambda^+-\lambda^-\right) \left(\lambda^+-\lambda^\times\right)} &
\frac{\hcs^2+\lambda^\times \lambda^-}{c^2\left(\lambda^+-\lambda^-\right) \left(\lambda^+-\lambda^\times\right)} & 
-\frac{\mu(u_d-u_g) (\beta-1) }{\left(\lambda^+-\lambda^-\right) \left(\lambda^+-\lambda^\times\right)}
\end{array}
\right).
\end{equation}
Just like for the eigenvalues, it can be easily verified that the
definitions (\ref{rightev:eq}) and (\ref{leftev:eq}) tend to the usual
expression for the left and right eigenvectors in the non stiff limit.
\subsubsection{Loss of Strict Hyperbolicity}

When either one of the following three cases occurs, $\beta\rightarrow
1,~\delta u\rightarrow 0,~ \mu\rightarrow 0$, (implying either
$\rho_g=0$ or $\rho_d=0$) the first, third and fourth right eigenvectors,
appear to become singular.
However, the asymptotic analysis shows that in these limits the
expressions, $\lambda^\pm \mp\hcs$, approach zero quadratically in
$(\beta-1)$ and linearly in both $\rho_d$ and $\du$, so that the first
and fourth right eigenvectors are always well defined.

While the third right eigenvector still diverges, the corresponding
left eigenvector tends to zero at the same rate (because
$\lambda^++\lambda^-$ is quadratic in $(\beta-1)$ and linear in both
$\rho_d$ and $\du$, whereas $\hcs^2+\lambda^+\lambda^-$, is quadratic
in both $(\beta-1)$ and $\du$, and linear in $\rho_d$), so that the
characteristic decomposition is non-singular in the above limits.
Note that this ``illness'' of the third right eigenvector could be
formally cured by renormalizing the third left and right eigenvectors
by the factor $f_3$ and $f_3^{-1}$, respectively, where
\begin{equation}
f_3=\frac{(\lambda^+-\lambda^\times)(\lambda^\times-\lambda^-)}{\mu\du(\beta-1)}.
\end{equation}

The eigensystem can still become singular when two eigenvalues
become identical, i.e. strict hyperbolicity is lost.
Inspection of Eq.~(\ref{lambda1:eq})-(\ref{lambda3:eq}) and
(\ref{phi:eq}) indicates that it is possible to have
$\lambda^-=\lambda^\times$ or $\lambda^+=\lambda^\times$ when $\phi=0$
or $\phi=\pi$, respectively. These relations are satisfied when
$\delta\hat u = \mp \hat c$, respectively, and simultaneously either
$\beta=1$, or $\mu=0$ is verified. As it appears from
(\ref{rightev:eq}) and (\ref{leftev:eq}), in this case the third
($\times$) eigenvectors (left and right) become parallel to either the
first ($-$) or fourth ($+$) eigenvectors, respectively.
When this is the case we modify our characteristic synthesis as 
suggested by Bell et al.~\cite{belletall1989}. For example, 
if $ \alpha_k=\sum_k l_k\cdot \delta W$ are the coefficients of 
the characteristic decomposition of a perturbation $\delta W$
with respect to the set of left eigenvalues $l_k$,
when $\lambda^\times-\lambda^\pm\leq \epsilon$, with typically,
$\epsilon\simeq 10^{-2}$, the above perturbation would be reconstructed as
\begin{gather}
\delta W = \sum_k \alpha_k r_k \rightarrow 
\alpha_\times r_\pm + \sum_{k\neq\times} \alpha_k r_k ,
\end{gather}
where $r_k$ is the set of right eigenvectors. 
\subsubsection{Addition of Endothermic Processes}

We can easily generalize the above results to include
stiff endothermic processes. Suppose the rate of 
change of the gas internal energy, $e=P/\rho_g(\gamma-1)$, is
\begin{equation}\label{dedt:eq}
\frac{de}{dt} = \Lambda(e,\rho_g).
\end{equation}
In this case the modified linear operator defined in Eq.~(\ref{aeff:eq})
becomes
\begin{equation}\label{aeffcool:eq}
 A^{\rm eff}_L =
\begin{pmatrix}
0 & \rho_g & 0 & 0 & 0 \\
0 & 0 & \frac{\kd+\beta\kg}{\ka\rho_g} & 0 & \frac{\kg(1-\beta)}{\ka} (u_d-u_g)\\
0 & \alpha\rho_g c^2-\frac{(1-\alpha)\Lambda_\rho\rho_g}{\Lambda_e} & 0 & 0  & 0 \\
0 & 0 & 0 & (u_d-u_g) & \rho_d \\
0 & 0 & \frac{\kd(1-\beta)}{\ka\rho_g} & 0 & \frac{\kg+\beta\kd}{\ka}(u_d-u_g)
\end{pmatrix},
\end{equation}
where
\begin{gather}\label{ceff_vars:eq}
\Lambda_e\equiv\frac{\partial\Lambda}{\partial e},
~~~\Lambda_\rho\equiv\frac{\partial\Lambda}{\partial \rho_g},~~~
\alpha(\Delta t)\equiv\frac{e^{\Lambda_e\Delta t }-1}{\Lambda_e\Delta t}, ~~0<\alpha<1.
\end{gather}
Then it can be shown that the above characteristic analysis remains valid
provided the gas sound speed is replaced
with the following {\it effective} value~\cite{mico07a}:
\begin{gather}\label{ceff:eq}
c_{\rm eff} = \left\{\left[1+\alpha(\gamma-1)-(1-\alpha)\frac{\Lambda_{\rho}\rho_g}{\Lambda_e e}\right] \frac{P}{\rho_g} \right\}^\frac{1}{2}.
\end{gather}
As discussed at length in~\cite{mico07a} the above scheme is applicable 
as long as the fluid is thermally stable~\cite{field65}, namely:
\begin{equation}
\frac{\Lambda_e e}{\Lambda_\rho \rho_g} > \frac{1-\alpha}{\alpha(\gamma-1)+1}.
\end{equation}

\subsection{Linearized Riemann Solver} \label{lrs:sec}
\begin{figure*}
\begin{center}
\includegraphics[width=0.4\textwidth,angle=90]{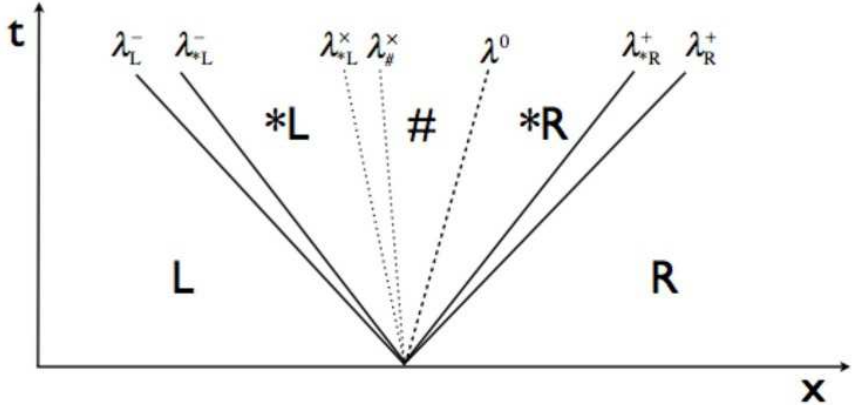}
\caption{Four-waves structure of the Riemann problem.
\label{rss:fig}}
\end{center}
\end{figure*}

The characteristic analysis derived in~\ref{charanalys:se} defines the
structure of the solution to the Riemann problem for the gas-dust
system. Unlike ordinary hydrodynamics, this is now characterized by
three acoustic wave families and one entropy wave propagating at the
fluid velocity\footnote{For the sake of clarity, in the following we
  continue to ignore the contact discontinuity in the dust component,
  which travels at speed $\lambda^d$ without interacting with the rest
  of the primitive variables.}.  The corresponding four characteristic
curves determine in general five distinct regions, corresponding to
the initial left (L) and right (R) states and three intermediate
states of the gas.  Fig.~\ref{rss:fig} illustrates this, for the case
in which $\lambda^\times<\lambda^0$.  In the $x-t$ plane, the solution
cone is separated by the unperturbed R and L states by the fastest and
slowest eigenvalues, respectively. The R-state is connected to a
$\ast$R-state by a rarefaction fan or a shock depending on whether
$\lambda^+_{R} $ is faster or slower than $\lambda^+_{*R} $.  The same
applies to the connection between the L-state and $\ast$L-state.  Then
if $\lambda^\times<\lambda^0$ as assumed in Fig.~\ref{rss:fig}, the
$\ast$R-state is connected to a $\#$-state by an ordinary contact
discontinuity, where the density jumps but pressure and velocity
remain constant.  Finally, another acoustic wave separates the
\#-state and $\ast$L-state, which again either takes the form of a
shock or a rarefaction wave depending on whether the characteristics
with speed $\lambda^\times_{\ast L}$ and $\lambda^\times_{\#} $
converge or diverge.

In order to calculate the intermediate states described above, we use
a linearized Riemann solver~\cite{whitham74}, which is discussed in
detail below.  Then, in general, given the input left and right states
$W_L, W_R$, we can use the left and right eigenvectors,
$\{l_k,r_k\}_{k = 1 \dots n}$ and the corresponding eigenvalues
$\lambda^k$, computed at some intermediate state, to decompose the
perturbation along the characteristics as follows:
\begin{gather*} 
\alpha^k = l^k \cdot (W_R - W_L) ,\\
W^k_R = W_L + \sum \limits_{k' = 1}^k \alpha_{k'} r_{k'} \hbox{ , } W^k_L = W^k_R - \alpha_k r_k  \hbox{ , } \lambda^k_{L,R} = \lambda^k(W^k_{L,R}).
\end{gather*}
Then the solution to the Riemann problem along the ray $x/t
= 0$,  $W_{RP}$,  is given by
\begin{gather*}
W_{RP} = W_L +
\sum \limits_{\max(\lambda^k_L , \lambda^k_R) < 0 } \alpha_k r_k +\sum \limits_{\lambda^k_L > 0 > \lambda^k_R, \lambda^k_L + \lambda^k_R < 0 } \alpha_k r_k + 
\sum \limits_{\lambda^k_L < 0 < \lambda^k_R} \frac{\lambda^k_L}{\lambda^k_L - \lambda^k_R} \alpha_k r_k ,
\end{gather*}
The first sum is the correction to the state for the waves that are
unambiguously have a negative speed; the second is a correction to the
state for transonic shocks; and the third a correction for transonic
rarefactions, that prevents the formation of entropy-violating shocks.
In the following we use a slight variation of this approach that is
better able to preserve reflection symmetries.

Recall that we have three acoustic-like wave families $k = +,-,\times$ 
with speeds $\lambda^k$ given in (\ref{lambda1:eq})-(\ref{lambda3:eq})
and the remaining wave propagates at the fluid speed, $\lambda^0$.
Assume for the moment that we can compute the eigenvalues
$\lambda^k$ at the intermediate states, $\ast$L, \#, $\ast$R,
following a procedure described in detail below.
Then the approximate Riemann solver is given as follows. 
\newline
\newline \noindent If $\lambda^0 > 0$, then \newline
\indent if $\min(\lambda^\times _{*L},\lambda^{\times} _\#) > 0$ or $\lambda^{\times} _{*L}>0>\lambda^{\times} _\#$ and $\lambda^{\times} _{*L}+\lambda^{\times} _\# >0 $
then
\begin{gather*}
W_{RP} = 
\begin{cases}
 W_{*L} \hbox{ if } \max(\lambda^-_L , \lambda^-_{*L}) < 0  \hbox{ or } \lambda^-_L > 0 > \lambda^-_{*L}, \lambda^-_L + \lambda^-_{*L} < 0 \\
W_L + \frac{\lambda^-_L}{\lambda^-_L - \lambda^-_{*L}} (W_{*L} - W_L)  \hbox{ if } \lambda^-_L < 0 < \lambda^-_{*L}\\
W_L  \hbox{ otherwise }
\end{cases}
\end{gather*}
\indent else if $\lambda^\times _{*L} <0< \lambda^\times _\#$ then 
\newline
\indent \indent $W_{RP} =  W_{*L} + \frac{\lambda^\times_{*L}}{\lambda^\times_{*L} - \lambda^\times_{\#}} (W_{\#} - W_{*L})$
\newline \indent else
\newline
\indent \indent$W_{RP} =  W_\#$
\newline \indent endif
\newline
\newline \noindent else if $\lambda^0< 0$, then \newline
\indent if $\max(\lambda^\times _{\#},\lambda^{\times} _{*R}) < 0$ or $\lambda^{\times} _{\#}>0>\lambda^{\times} _{*R}$ and $\lambda^{\times} _{\#}+\lambda^{\times} _{*R} <0 $
then
\begin{gather*}
W_{RP} = 
\begin{cases}
 W_{*R} \hbox{ if } \max(\lambda^+_{*R} , \lambda^+_{R}) > 0  \hbox{ or } \lambda^+_{*R} > 0 > \lambda^+_{R}, \lambda^+_{*R} + \lambda^+_{R} > 0 \\
W_R + \frac{\lambda^+_R}{\lambda^+_R - \lambda^+_{*R}} (W_{*R} - W_R)  \hbox{ if } \lambda^+_{*R} < 0 < \lambda^+_{R}\\
W_R  \hbox{ otherwise }
\end{cases}
\end{gather*}
\indent else if $\lambda^\times _\# <0< \lambda^\times _{*R}$ then 
\newline
\indent \indent $W_{RP} =  W_{*R} + \frac{\lambda^\times_{*R}}{\lambda^\times_{*R} - \lambda^\times_{\#}} (W_{\#} - W_{*R})$
\newline \indent else
\newline
\indent \indent$W_{RP} =  W_\#$
\newline \indent endif
\newline endif

In order to complete our description of the approximate Riemann
solver, we now outline the method for computing the intermediate
states.  Basically we integrate the characteristic equations along the
characteristic directions to relate the jumps experienced by the
primitive variables across adjacent states~\cite{toro97}.  The
characteristic directions are assumed to be constant and are computed
at the foot of the characteristics.  The characteristic equations are
obtained by setting $l^k\cdot dW=0$ for each eigenstate and they read
\begin{eqnarray}
dp -\chi^-_g(\lambda^\times,\lambda^+) du_g - \chi^-_d(\lambda^\times,\lambda^+)  du_d = 0, \\
dp -c^2 d\rho_g = 0 ,\\
dp -\chi^\times_g(\lambda^-,\lambda^+)  du_g - \chi_d^\times(\lambda^-,\lambda^+)   du_d = 0, \\
dp -\chi^+_g(\lambda^\times,\lambda^-)  du_g - \chi_d^+(\lambda^\times,\lambda^-)  du_d = 0,
\end{eqnarray}
where the characteristic slopes in phase-space are given by
\begin{eqnarray} \label{chig:eq}
\chi^+_g(\lambda^\times,\lambda^-) &=& \frac{\lambda^-+\lambda^\times}{\hcs^2+\lambda^-\lambda^\times} \rho_gc^2, \\ \label{chid:eq}
\chi^+_d(\lambda^\times,\lambda^-) &=& \frac{\mu \delta u(-1+\beta)}{\hcs^2+\lambda^-\lambda^\times} c^2,
\end{eqnarray}
and the analogous expressions for $\chi_{g,d}^-,\chi_{g,d}^\times$ are 
obtained by permutation of the symbols $+,-,\times$ in the above equations.
The above procedure of integrating the characteristic equations yields
12 equations in 12 variables.  However, neither the gas or dust fluid
velocities nor the gas pressure change across the contact
discontinuity, reducing the system to 9 equations in the following 9
variables:
$\rho_{g,*L},u_{g,*L},P_{g,*L},u_{d,*L},\rho_{g,*R},u_{g,*R},P_{g,*R},u_{d,*R},\rho_{g,\#}$,
with $u_{g,\#},P_{g,\#},u_{d,\#}$ coinciding with the corresponding variable
of either the *L or *R state.  In order to keep the system linear we
make two further approximations when computing the characteristic
directions~(\ref{chig:eq})-(\ref{chid:eq}) in the intermediate states:
(1) we assume that the gas-dust velocity difference, $u_g-u_d$, in the
intermediate *L,*R-states, can be approximated with the values
corresponding to the L,R-states, respectively.  And (2), that, when
connecting the \#-state with the *L-state (*R-state) via the $-~(+)$
characteristic, i.e. when $\lambda^\times<\lambda^0$
($\lambda^\times>\lambda^0$), we can use the density of the *L-state
(*R-state) instead of the value corresponding to the \#-state.  With
this latter approximation the relations between velocity jumps across
the characteristic line $\times$, remain unchanged whether
$\lambda^\times<\lambda^0$ or $\lambda^\times>\lambda^0$.  The
solution then reads
\begin{eqnarray} \label{rsug:eq}
u_{g,*R} &= &u_{g,R} +   \nonumber \\ & &\frac{(P_L-P_R)(1-\frac{\etaL}{\etas})+(u_{g,L}-u_{g,R})(\ttal+\frac{\ttas}{\etas}\etaL)+(u_{d,L}-u_{d,R})\frac{\ttas+\ttal}{\etas}}{(\ttal-\ttar)-\etar\frac{\ttal+\ttas}{\etas}+\etaL\frac{\ttar+\ttas}{\etas}}, \\
P_{*R}& =& P_{R} + \ttar(u_{g,R}-u_{g,*R}), \\
u_{d,*R} &=& u_{d,R} + \etar (u_{g,R}-u_{g,*R}), \\
\rho_{g,*R}& =& \rho_{g,R} + \frac{P_{*R}-P_R}{c_R^2}, \\
u_{g,*L} &=& u_{g,L} + \nonumber \\ & & \frac{(P_L-P_R)(1-\frac{\etar}{\etas})+(u_{g,L}-u_{g,R})
(\ttar+\frac{\ttas}{\etas}\etar)+(u_{d,L}-u_{d,R})\frac{\ttas+\ttar}{\etas}}
{(\ttal-\ttar)-\etar\frac{\ttal+\ttas}{\etas}+\etaL\frac{\ttar+\ttas}{\etas}} \\
P_{*L} &=& P_{L} + \ttal(u_{g,L}-u_{g,*L}), \\
u_{g,*L} &=& u_{g,L} + \etaL(u_{g,L}-u_{g,*L}), \\
\rho_{g,*L}& =& \rho_{g,L} + \frac{P_{*L}-P_L}{c_L^2}, \\
\rho_{g,\#} &= &
\begin{cases}
\rho_{g,*R} + \frac{P_{*L}-P_{*R}}{c^2_{*R}} \hbox{\hskip 1cm if} \lambda^\times_{*L}>\lambda^0, \\
\rho_{g,*L} + \frac{P_{*R}-P_{*L}}{c^2_{*L}} \hbox{\hskip 1cm if} \lambda^\times_{*R}<\lambda^0,
\end{cases} \\  \label{rsrd:eq}
\rho_{d,*L} &=& \rho_{d,*R}=\rho_{d,\#}=\frac{\rho_{d,L}+\rho_{d,R}}{2},
\end{eqnarray}
where we have defined the following coefficients,
\begin{gather} \label{tetas:eq}
\ttar =\frac{\chi^-_{g,R}\chi^\times_{d,R} - \chi^\times_{g,R}\chi^-_{d,R}}{\chi^-_{d,R}-\chi^\times_{d,R}} = -\left(\frac{\rho_{g}c^2}{\lambda^+}\right)_R,~
\ttal =\frac{\chi^+_{g,L}\chi^\times_{d,L} - \chi^\times_{g,L}\chi^+_{d,L}}{\chi^+_{d,L}-\chi^\times_{d,L}}= -\left(\frac{\rho_{g}c^2}{\lambda^-}\right)_L, \\ 
\etar = \frac{\chi^-_{g,R}-\chi^\times_{g,R}}{\chi^-_{d,R}-\chi^\times_{d,R}} = 
\left[\frac{\rho_{g}}{\lambda^+}\frac{(\lambda^+)^2  -\hcs^2}{\mu\du(\beta-1)}\right]_R,
~~\etaL = \frac{\chi^+_{g,L}-\chi^\times_{g,L}}{\chi^+_{d,L}-\chi^\times_{d,L}} =
\left[\frac{\rho_{g}}{\lambda^-}\frac{(\lambda^-)^2  -\hcs^2}{\mu\du(\beta-1)}\right]_L,\\
\ttas = \frac{\chi^-_{g,*R}\chi^+_{d,*L} - \chi^+_{g,*L}\chi^-_{d,*R}}{\chi^+_{d,*L}-\chi^-_{d,*R}}, 
~~\etas = \frac{\chi^+_{g,*L}-\chi^-_{g,*R}}{\chi^+_{d,*L}-\chi^-_{d,*R}} \label{etas:eq}.
\end{gather}
The extra symbols L,*L,*R,R in the subscript of the $\chi$ functions
defined in~(\ref{chig:eq})-(\ref{chid:eq}) indicate the state at which
the characteristic directions is defined.  Note that while the
definition of the coefficients $\ttal,\ttar,\etaL,\etar$ is based on
the known L,R states, computing $\ttas$ and $\etas$ involves knowledge
of the intermediate states, *L or *R. It was in order to avoid the
nonlinearities arising from such dependencies that we have introduced
the above approximations (1) and (2).
As a sanity check we may note that the non-stiff limit for the
characteristic slopes in phase space read
\begin{gather}
\lim_{\beta\to 1} \chi^{\pm}_g = \mp \rho c,~~~\lim_{\beta\to 1} \chi^\pm_d, \chi^\times_g = 0, ~~~\lim_{\beta\to 1} \chi^\times_d  =  \infty ,
\end{gather}
so that the coefficients in (\ref{tetas:eq})-(\ref{etas:eq}) tend to
\begin{gather}
\lim_{\beta\to1} \theta_{L} = \rho_{L}c_{L}, ~~\lim_{\beta\to1} \theta_{R} = -\rho_{R}c_{R},
~~\lim_{\beta\to1} \ttas = \hbox{const.,}~~
\lim_{\beta\to1} \eta_{L,R} = 0, ~~\lim_{\beta\to1} \etas =  \infty,
\end{gather}
and the Riemann solver solutions (\ref{rsug:eq})-(\ref{rsrd:eq}) reduce 
to the usual expressions~\cite{toro97}.

\subsection{Godunov Predictor in One Dimension}

With the operator $A^{\rm eff}$ and the sets of left and right
eigenvectors that we have worked out in section~\ref{charanalys:se}, the
Godunov predictor step is carried out as usual as follows.
First the local slopes are defined. In particular at each point left
and right one-sided slopes as well as cell centered slopes are
evaluated and then a final choice on the local slope $\Delta W_{i}$
is defined by using van Leer limiter.
The upwind, time averaged left $(-)$ and right $(+)$ states at
cell interfaces due to fluxes in the normal direction, $q$, are then
reconstructed as:

\begin{equation}
W_{i,\pm,q}=W^n_i+\frac{1}{2}\left(I-\frac{\Delta t}{\Delta x}A^{\rm eff}_{i}\right) P_{\pm}(\Delta W_{i}),
\end{equation}
where
\begin{equation}
P_{\pm}(W)=\sum_{\pm\lambda_k>0}(l_k\cdot W)\cdot r_k.
\end{equation}
The source term component is likewise accounted for as
\begin{equation} \label{wpm:eq}
W_{i,\pm,q}=W_{i,\pm,q}+\frac{\Delta t}{2}  {\mathcal I}^0_{K}(\Delta t/2) S.
\end{equation}

The fluxes at the cell faces $F_{i + \frac{1}{2}}$ are
computed by solving the Riemann problem with left and right states
given by $\left(W_{i,+} , W_{i+1,-}\right)$ to obtain
$W^{n+\frac{1}{2}}_{i+\frac{1}{2}}$ and computing $F_{i+\frac{1}{2}} =
F\left(W^{n+\frac{1}{2}}_{i+\frac{1}{2}}\right)$.
To modify this procedure to account for the effective dynamics, we use
the characteristic analysis of the effective dynamics to perform each
of the three steps. The projection operator and any limiting in
characteristic variables is done using the eigenvectors and
eigenvalues for the effective dynamics derived in Sec.
\ref{charanalys:se}.  Finally, the approximate Riemann solvers we use
was defined in~\ref{lrs:sec}.

\subsection{Extension to More than One Dimension}

For directionally unsplit schemes in $D$ dimensions an
additional step is required in order to correct the time-averaged
left/right states at cell interfaces, $W_{i,\pm,d}$ in
Eq.~(\ref{wpm:eq}), for the effects of $D-1$ fluxes perpendicular to the
cell interface normal direction. Based on Eq.~(\ref{dsol:eq})
the effect of the stiff source term would be accounted for by
carrying out for each additional direction, $q$, a transformation
\begin{equation} \label{ad:eq}
A_{q}\rightarrow  {\mathcal I}^0_{K }(\Delta t/2) 
A_{L,q}+u_q{\rm I} \equiv A^{\rm eff}_{q},
\end{equation}
analogous to that described in Eq.~(\ref{aeff:eq}).
In the method proposed by \cite{colella90,saltzman94} the corrections
due to transverse fluxes are computed according to a conservative
scheme. For example in two dimensions\footnote{ Notation in Eq.
  (\ref{consupdate:eq}) indicates that primitive variables are
  converted into conservative variables which are then updated through
  conservative fluxes and then converted back into primitive form.},
with $q=x$
\begin{gather} \label{consupdate:eq}
W_{{i,j},\pm,x} = W_{{i,j},\pm,x} - \frac{\Delta t}{2 \Delta y} \nabla_U W \;
\left( F^y_{{i,j+\frac{1}{2}}} - F^y_{{i,j-\frac{1}{2}}} \right),
\end{gather}
where the input $W_{{i,j},\pm,x}$ is computed using a one-dimensional
Godunov calculation as in the previous section, as are the fluxes
$F^y_{{i,j+\frac{1}{2}}}$.  
The above transformations imply the following transverse flux
corrections for the gas and dust velocity, $\delta u_g,~\delta
u_v$,respectively:

\begin{gather}
\delta u_{g,x} \rightarrow \delta u_{g,x} +\frac{\Delta t}{2\Delta y}
\frac{\kg}{\ka} (1-\beta) 
\left[
\frac{p_{i,j+\frac{1}{2}} - p_{i,j-\frac{1}{2}}}{\rho_g}-
(u_d-u_g) \left(u_{y,i,j+\frac{1}{2}}  -u_{y,i,j-\frac{1}{2}}\right) \right],\\
\delta u_{d,x} \rightarrow \delta u_{d,x} -\frac{\Delta t}{2\Delta y}
\frac{\kd}{\ka} (1-\beta) 
\left[
\frac{p_{i,j+\frac{1}{2}} - p_{i,j-\frac{1}{2}}}{\rho_g}-
(u_d-u_g) \left(u_{y,i,j+\frac{1}{2}}  -u_{y,i,j-\frac{1}{2}}\right) \right],
\end{gather}
where, unless explicitly indicated, all quantities are evaluated at 
cell center, $i,j$.
\section{Stability Considerations} \label{sa:se}
The stability properties of the above modified Godunov's method are analogous 
to those discussed in~\cite{mico07a} when considering endothermic source terms.
In particular we note that the inspection of the characteristic analysis shows that
the {\it sub-characteristic condition} for the characteristic speeds at equilibrium 
is always satisfied. 
This condition, while being necessary for the stability of our
linearized system \cite{whitham74}, also guarantees that the numerical
solution tends to the solution of the equilibrium equation as the
relaxation time tends to zero \cite{chleli94}.  

In addition, since
the structure of the equations and the numerical framework, including
the Riemann solver, remain fundamentally unaltered with respect to classic
Godunov's schemes,  we expect the usual stability conditions to apply,
namely the familiar CFL condition on the time step
\begin{equation} \label{cfl:eq}
\max (| \lambda_* |) \frac{\Delta t}{\Delta x} \le 1, \hspace{1 cm} *=-,0,\times,d,+.
\end{equation}
As for the predictor corrector method described in Sec.~\ref{sim:se}, 
A-stability properties the $\alpha-$QSS method have been discussed in detail 
by Mott et al.~\cite{Mottetal2000}. Although not specifically mentioned by the
authors, the same stability analysis they present shows immediately 
that the $\alpha-$QSS method is also L-stable.

\section{Tests} \label{tests:se}
\subsection{Particle Scheme} \label{convrate_ps:se}
\begin{table}  
\caption{Particle Scheme Convergence Rates.\label{pscrA.tab}}
\begin{tabular*}{\textwidth}{@{\extracolsep{\fill}}lccccc}
\hline
\hline
N$_{\rm steps}$ & $E_x$ & $R_x$ & $E_v$ & $R_v$ & $\kappa_0\Delta t$ \cr
\hline
2 & 4.6E-3 & 1.4 & 8.1E-4 & 0.1 & 5.0E1\cr
4 & 1.7E-3 & 2.2 & 7.8E-4 & 0.3 & 2.5E1\cr
8 & 3.8E-4 & 2.2 & 6.5E-4 & 2.7 & 1.2E1\cr
16 & 8.4E-5 & 2.0 & 1.0E-4 & 2.0 & 6.2E0\cr
32 & 2.1E-5 & 2.0 & 2.5E-5 & 0.9 & 3.1E0\cr
64 & 5.2E-6 & 2.0 & 1.4E-5 & 1.8 & 1.6E0\cr
128 & 1.3E-6 & 2.0 & 4.0E-6 & 1.9 & 7.8E-1\cr
256 & 3.2E-7 & 2.0 & 1.0E-6 & 2.0 & 3.9E-1\cr
512 & 8.1E-8 & 2.0 & 2.6E-7 & 2.0 & 2.0E-1\cr
1024 & 2.0E-8 & 2.0 & 6.6E-8 & 2.0 & 9.8E-2\cr
2048 & 5.0E-9 & 2.0 & 1.7E-8 & 2.0 & 4.9E-2\cr
\hline
\hline
\end{tabular*}
\end{table}

\begin{table}  
\caption{Particle Scheme Convergence Rates.\label{pscrB.tab}}
\begin{tabular*}{\textwidth}{@{\extracolsep{\fill}}lccccc}
\hline
\hline 
N$_{\rm steps}$ & $E_x$ & $R_x$ & $E_v$ & $R_v$ & $\kappa_0\Delta t$ \cr
\hline 
2 & 4.4E-3 & 1.3 & 9.4E-4 & 1.7 & 5.0E5\cr
4 & 1.7E-3 & 2.2 & 2.9E-4 & 3.2 & 2.5E5\cr
8 & 3.8E-4 & 2.2 & 3.1E-5 & 0.5 & 1.2E5\cr
16 & 8.2E-5 & 2.0 & 2.2E-5 & 2.0 & 6.2E4\cr
32 & 2.0E-5 & 2.0 & 5.5E-6 & 2.1 & 3.1E4\cr
64 & 5.1E-6 & 2.0 & 1.2E-6 & 2.1 & 1.6E4\cr
128 & 1.3E-6 & 2.0 & 2.8E-7 & 2.1 & 7.8E3\cr
256 & 3.2E-7 & 2.0 & 6.6E-8 & 2.1 & 3.9E3\cr
512 & 8.0E-8 & 2.0 & 1.6E-8 & 2.1 & 2.0E3\cr
1024 & 2.0E-8 & 2.0 & 3.6E-9 & 2.3 & 9.8E2\cr
2048 & 5.0E-9 & 2.0 & 7.4E-10 & 2.7 & 4.9E2\cr
\hline
\hline
\end{tabular*}
\end{table}
Before testing the convergence of the gas-dust scheme, we investigate
the performance of the particle scheme presented in Sec.~\ref{pis:se}.
For the purpose we consider an individual particle propagating through
a medium with specified density, velocity and pressure distributions.
As expected~\cite{Mottetal2000}, our scheme reproduces the {\it exact}
solution when the particle moves through a uniform medium, and when
either the coupling constant or the background fluid has linear time
dependence.  Thus in the following for our testing purposes we use a
non-linear time dependent density, velocity and pressure field
corresponding to a one-dimensional, right propagating sound wave
solution.  Note that, although not presented here, we obtain
equivalent convergence rates as shown below when an external force
acting on the particle is included.  The wave quantities denoted with
a `w' subscript are given by~\cite{lali6}
\begin{gather}
u_w(x,t) = u_{w,0} +B\,\sin\left\{x-u_{w,0}t- \left[c_0+\frac{\gamma+1}{2}(u_w-u_{w,0})\right] t\right\},  \\
\rho_w(x,t) = \rho_{w,0}\left[ 1+\frac{\gamma-1}{2}\frac{u_w}{c_0}\right]^\frac{2}{\gamma-1},\quad  
P_w(x,t) = P_{w,0}\left[ 1+\frac{\gamma-1}{2}\frac{u_w}{c_0}\right]^\frac{2\gamma}{\gamma-1},
\end{gather}
where, $c_0=\sqrt{\frac{\gamma P_{w,0}}{\rho_{w,0}}}$, and we use
\begin{gather}
B=10^{-2},\quad u_{w,0}=2,\quad \rho_{w,0}=1,\quad P_{w,0}=1,\quad\gamma=\frac{5}{3}.
\end{gather}
Using the scheme described in Sec.~\ref{pis:se} we advance the
particle position and velocity to a solution time $t_{end}$ using a
progressively larger number of steps, $N_{steps}$, and a
correspondingly smaller time-step, $\Delta t_N=t_{end}/N_{steps}$.  We
then estimate the error by comparing solutions with resolution
different by a factor 2 and use Richardson extrapolation to measure
the convergence rates. So, if $s(m)$ is the solution obtained
using $m$ time-steps, the convergence rate is given by
\begin{equation}
R(m) = \frac{\ln\left|\frac{s(m)-s(2m)}{s(2m)-s(4m)}\right|}{\ln(2)}.
\end{equation}
In Table~\ref{pscrA.tab} we report results for a choice of the coupling
constant $\kappa_0$ such that the propagation regime goes from mildly
stiff to non-stiff.  From left to right the table columns 
report the number of steps, the error
and convergence rate of the particle position and velocity and, in the
last column, the stiffness parameter, $\kappa_0\Delta t$.  The scheme
is clearly second order accurate over the stiffness range reported in
the Table, even though the convergence rate of the velocity appears to
be slower for small $N_{steps}$. However, this is not due to a
decrease in convergence rate in the stiff regime.  As illustrated in
Table~\ref{pscrB.tab}, the particle scheme is still second order
accurate even for much larger values of the stiffness parameter.

\subsection{Convergence Rates in Smooth Flows} \label{convrate:se}
In this section we test the convergence of the method presented in
this paper, by studying the propagation of small perturbations in a
gas-dust system with a long and short drag relaxation timescale.  
We consider both a two-fluid method in which the dust is
modeled as a fluid (and dust particles are not used), as well as the
full fluid+particle method.  In our test, the fluid component is
initialized with uniform density, pressure and velocity values, except
for a sinusoidal perturbation with small amplitude superposed to the
x-velocity component. In formulae
\begin{gather}
u_{g,x} =  u_0\, [1+A\cos(2\pi\,{\bf k\cdot r}+\pi) ], \label{vginit:eq}\\
\rho  = \rho_0  =\gamma = 1.4,\quad p=p_0   =  0.5,\quad u_y = u_{y0} = 0.7,
\quad u_0=0.5,\label{vinit0:eq}
\end{gather}
where ${\bf r}$ is the position vector. Similarly the dust particles
are uniformly distributed on the grid in order to produce a uniform
density, and their velocity is initialized as 
\begin{gather}
u_{d,x} =  u_0\, [1+A\cos(2\pi\,{\bf k\cdot r}) ], \label{vdinit:eq}
\end{gather}
that is half a period out of phase with respect to $u_{g,x}$. When
testing the hybrid method in two dimensions we use 4 particles per
cell and a we interpolate the particle quantities to the grid using a
Triangular-Shape-Cloud interpolation scheme~\cite{mico07b}.

The above initial conditions produce a sinusoidal wave with amplitude ${A}$
propagating in the domain along the direction defined by the vector
${\bf k}$.  While we have experimented with various values for the parameters 
${A}$, ${\bf k}$ and $\kappa_0$, below we present results for a few cases only,
summarized in Table~\ref{runset.tab}. 
\begin{table}  
\caption{Run Set\label{runset.tab}}
\begin{tabular*}{\textwidth}{@{\extracolsep{\fill}}lcccccc}
\hline \hline 
run & $D$ & ${A}$ & ${\bf k}$ & $\kappa_0$ &$\rho_d/\rho_g$ & Note$^\dagger$ \cr
\hline
A& 1& $1.4\times 10^{-2}$ & $(1,0)$  & $ 1$    & 1 & {\rm two-fluid} \cr
B& 1& $1.4\times 10^{-2}$ & $(1,0)$  & $10^6$  & 1 & {\rm two-fluid} \cr  
C& 1& $1.4\times 10^{-2}$ & $(1,0)$  & $10^{6}$& $10^{-3} $&{\rm two-fluid}\cr
D& 1& $1.4\times 10^{-2}$ & $(1,0)$  & $10^{6}$& $10^{3}$&{\rm two-fluid}\cr 
E& 1& $1.4\times 10^{-2}$ & $(1,0)$  & $2\times 10^{2}$& $1$&{\rm two-fluid}\cr 
F& 2& $1.4\times 10^{-2}$ & $(2/\sqrt{5},1/\sqrt{5})$ & $1 $    & 1 & {\rm hybrid} \cr 
G& 2& $1.4\times 10^{-2}$ & $(2/\sqrt{5},1/\sqrt{5})$ & $10^6$  & 1 & {\rm hybrid} \cr 
H& 2& $1.4\times 10^{-2}$ & $(2/\sqrt{5},1/\sqrt{5})$ & $10^6$  & $10^{-3}$&{\rm hybrid} \cr 
I& 2& $1.4\times 10^{-2}$ & $(2/\sqrt{5},1/\sqrt{5})$ & $10^{6}$& $10^{3}$&{\rm hybrid} \cr 
L& 2& $1.4\times 10^{-2}$ & $(2/\sqrt{5},1/\sqrt{5})$ & $10^{2}$& $1$&{\rm hybrid} \cr 
\hline
\end{tabular*}
\qquad\llap{$^\dagger$} {\it hybrid}=hydro + dust particles.
\end{table}
In particular we consider a perturbation amplitude $A=1.4\times
10^{-2}$ and adopt coupling coefficients
$\kappa_0=1,10^2,~10^6$ to explore the non-stiff and stiff regimes
respectively and, for simplicity, we report only results from
one dimensional tests for the two-fluid model and from two-dimensional 
tests for the hybrid model.

In order to measure the rate at which the numerical solution converges,
for each problem we carry out a set of 5 simulation runs employing 
$N_{cell}=16,32,64,128,256$ for a total range of 32.  Note
that the stiffness conditions do not change significantly as the grid
is refined within the range of resolutions considered here. 
The convergence rate is measured using Richardson extrapolation.
Given the numerical solution $q_{r}$ at a given resolution $r$ 
we first estimate the error at a given point $(i,j)$, as
\begin{equation} \label{numerr:eq}
\varepsilon_{r;i,j} = q_{r}(i,j) - \bar q_{r+1}(i,j),
\end{equation}
where $\bar q_{r+1}$ is the solution at the next finer resolution,
properly spatially averaged onto the coarser grid.
We then take the n-norm of the error
\begin{equation} \label{lnorm_n:eq}
L_n = \| \varepsilon_{r} \|_n =  \left( \sum |\varepsilon_{r;i,j}|^n  v_{i,j}\right)^{1/n},
\end{equation}
where, $v_{i,j}=\Delta x^2$ is the cell volume, 
and estimate the convergence rate as
\begin{equation}
R_n = \frac{ \ln[L_n(\varepsilon_r)/L_n(\varepsilon_s)] }{ \ln (\Delta x_r / \Delta x_s) }.
\end{equation}
For each studied case listed in Table~\ref{runset.tab}, we produce a
corresponding Table~\ref{caseA.tab}-\ref{caseD.tab} reporting the
$L_1,~L_2$ and $L_\infty$ norms of the error and the corresponding
convergence rates, $R_1,~R_2$ and $R_\infty$, as defined above.  Note
that the convergence rate of the dust component is evaluated in a way
analogous to the fluid components, i.e. by considering the convergence
of the error of the fluid representation of the particles.

\subsubsection{Two-fluid} \label{tfconvrate:se}

Cases A-E test the performance of the two-fluid scheme in which the
dust is fully treated as a fluid using the scheme described in
Sec.~\ref{sdc:se}--\ref{tfps:se} and the particle scheme is not used.
The five tests correspond to the following cases: (A) non-stiff,
(B-D) stiff but with a range of dust-to-gas ratios and (E) mildly
stiff case.  In all tests the perturbation is along the x-axis, ${\bf
  k}=(1,0)$, although we have tested that the performance is the same
in more than one dimension.  The convergence rates for each case are
reported in Tables~(\ref{caseA.tab})-(\ref{caseE.tab}), respectively,
for the density and velocity of the gas and dust. Errors in the gas
pressure are not reported but exhibit the same behavior.  In the
non-stiff case (A) all quantities converge with second order accuracy,
except the dust density, which is somewhat expected since in the
absence of drag a pressure-less fluid becomes singular.  In the stiff
cases (B-D), the scheme produces second order accurate results
independent of the dust-to-gas ratio. In the intermediate regime,
$\kappa_0\Delta t\sim 1$, the convergence rate is between first and
second order accurate as expected theoretically~\cite{mico07a}. Thus,
the two-fluid algorithm for the dust-fluid system is second
order accurate in both the stiff and non-stiff regimes, and
somewhat less accurate in between.

\subsubsection{Full Scheme} \label{fsconvrate:se} 

Cases F-L test the performance of the full fluid+particle
in which the dust is represented by a
set of particles whose velocity and position are updated with the
scheme described in Sec.~\ref{pis:se}.
As in the previous section,
the five tests correspond to the following cases: (F) non-stiff,
(G-I) stiff but with a range of dust-to-gas ratios and (L) mildly
stiff case. In all tests the perturbation is skewed with respect 
to the x-axis, ${\bf k}=(2,1)/\sqrt{5}$.
The convergence rates for each case are
reported in Tables~(\ref{caseF.tab})-(\ref{caseL.tab}), respectively,
for all of the gas and dust variables.

Inspection of the reported tables shows that, as expected, in the
non-stiff regime (Table~\ref{caseF.tab}) the error drops with second
order accuracy, this time even for the dust density.  In the stiff
regime and, unless the fluid mass dominates over the dust mass, the
error on the fluid quantities drops approximately with first order
accuracy, see Table~\ref{caseG.tab}-\ref{caseI.tab}.  Since, as
illustrated in the Sec.~\ref{convrate_ps:se} and~\ref{tfconvrate:se},
both the particle scheme and the two-fluid scheme retain their second
order accuracy irrespective of the stiffness conditions, the worsening
in accuracy is most likely due to the difficulty of coupling the gas
and the dust fully self-consistently in the stiff regimes.  However,
the scheme is stable and convergent, though only first order, which
would not have been trivially expected.  Finally, in the mildly stiff
regime, the convergence rate reduces to approximately first order, as
shown in Tables~\ref{caseL.tab} dust-to-gas ratio of 1.  This is
however, not unexpected because in this case accuracy of the two-fluid
algorithm also drops.

\begin{table}  
\begin{small}
\caption{Convergence Rates: Case : $A=1.4\times 10^{-2},~{\bf k}=(1,0),~\rho_d/\rho_g=1$.\label{caseA.tab}}
\begin{tabular*}{\textwidth}{@{\extracolsep{\fill}}lcccccccccccc}
\hline
\hline \cr
N$_{\rm cells}$ & $L_1$ & $R_1$ & $L_2$ & $R_2$ & $L_\infty$ & $R_\infty$ & $L_1$ & $R_1$ & $L_2$ & $R_2$ & $L_\infty$ & $R_\infty$ \cr \cline{2-13} \cr
& \multicolumn{6}{l}{\bf density-gas} & \multicolumn{6}{l}{\bf x-vel-gas} \cr
   16 &    6.4E-07 &    --  &    1.4E-06 &    --  &    4.1E-06 &    --  &    3.2E-06 &    --  &    7.1E-06 &    --  &    2.0E-05 &    --  \cr 
   32 &    1.6E-07 &    2.0 &    3.6E-07 &    2.0 &    1.0E-06 &    2.0 &    8.1E-07 &    2.0 &    1.8E-06 &    2.0 &    5.1E-06 &    2.0 \cr 
   64 &    4.0E-08 &    2.0 &    8.8E-08 &    2.0 &    2.5E-07 &    2.0 &    2.0E-07 &    2.0 &    4.5E-07 &    2.0 &    1.3E-06 &    2.0 \cr 
  128 &    9.9E-09 &    2.0 &    2.2E-08 &    2.0 &    6.3E-08 &    2.0 &    5.1E-08 &    2.0 &    1.1E-07 &    2.0 &    3.2E-07 &    2.0 \cr
\cline{2-7} \cline{8-13} \cr
& \multicolumn{6}{l}{\bf density-dust} & \multicolumn{6}{l}{\bf x-vel-dust} \cr
   16 &    1.3E-05 &    --  &    2.9E-05 &    --  &    8.2E-05 &    --  &    4.6E-06 &    --  &    1.0E-05 &    --  &    2.9E-05 &    --  \cr 
   32 &    6.1E-06 &    1.1 &    1.4E-05 &    1.1 &    3.8E-05 &    1.1 &    1.2E-06 &    2.0 &    2.6E-06 &    2.0 &    7.2E-06 &    2.0 \cr 
   64 &    2.9E-06 &    1.0 &    6.5E-06 &    1.0 &    1.9E-05 &    1.0 &    2.9E-07 &    2.0 &    6.4E-07 &    2.0 &    1.8E-06 &    2.0 \cr 
  128 &    1.5E-06 &    1.0 &    3.2E-06 &    1.0 &    9.2E-06 &    1.0 &    7.2E-08 &    2.0 &    1.6E-07 &    2.0 &    4.5E-07 &    2.0 \cr 
\hline
\hline
\end{tabular*}
\end{small}
\end{table}

\begin{table}  
\begin{small}
\caption{Convergence Rates: Case : $B=1.4\times 10^{-2},~{\bf k}=(1,0),~\rho_d/\rho_g=1 $.\label{caseB.tab}}
\begin{tabular*}{\textwidth}{@{\extracolsep{\fill}}lcccccccccccc}
\hline
\hline \cr
N$_{\rm cells}$ & $L_1$ & $R_1$ & $L_2$ & $R_2$ & $L_\infty$ & $R_\infty$ & $L_1$ & $R_1$ & $L_2$ & $R_2$ & $L_\infty$ & $R_\infty$ \cr \cline{2-13} \cr
& \multicolumn{6}{l}{\bf density-gas} & \multicolumn{6}{l}{\bf x-vel-gas} \cr
   16 &    3.2E-06 &    --  &    7.1E-06 &    --  &    2.1E-05 &    --  &    9.3E-07 &    --  &    2.3E-06 &    --  &    9.6E-06 &    --  \cr 
   32 &    8.0E-07 &    2.0 &    1.8E-06 &    2.0 &    5.2E-06 &    2.0 &    2.1E-07 &    2.1 &    4.9E-07 &    2.2 &    2.3E-06 &    2.1 \cr 
   64 &    2.0E-07 &    2.0 &    4.4E-07 &    2.0 &    1.3E-06 &    2.1 &    5.7E-08 &    1.9 &    1.3E-07 &    1.9 &    4.5E-07 &    2.3 \cr 
  128 &    4.9E-08 &    2.0 &    1.1E-07 &    2.0 &    3.1E-07 &    2.0 &    1.6E-08 &    1.8 &    3.6E-08 &    1.8 &    1.2E-07 &    1.9 \cr
\cline{2-7} \cline{8-13} \cr
& \multicolumn{6}{l}{\bf density-dust} & \multicolumn{6}{l}{\bf x-vel-dust} \cr
   16 &    3.7E-05 &    --  &    8.9E-05 &    --  &    4.0E-04 &    --  &    1.3E-06 &    --  &    2.9E-06 &    --  &    9.1E-06 &    --  \cr 
   32 &    1.0E-05 &    1.9 &    2.5E-05 &    1.8 &    1.3E-04 &    1.6 &    3.4E-07 &    1.9 &    7.6E-07 &    1.9 &    3.4E-06 &    1.4 \cr 
   64 &    2.6E-06 &    2.0 &    6.0E-06 &    2.1 &    2.0E-05 &    2.7 &    7.3E-08 &    2.2 &    1.7E-07 &    2.2 &    7.3E-07 &    2.2 \cr 
  128 &    7.7E-07 &    1.8 &    1.7E-06 &    1.8 &    4.8E-06 &    2.1 &    1.8E-08 &    2.0 &    4.1E-08 &    2.0 &    1.9E-07 &    1.9 \cr 
\hline
\hline
\end{tabular*}
\end{small}
\end{table}

\begin{table}  
\begin{small}
\caption{Convergence Rates: Case : $C=1.4\times 10^{-2},~{\bf k}=(1,0),~\rho_d/\rho_g=10^{-3} $.\label{caseC.tab}}
\begin{tabular*}{\textwidth}{@{\extracolsep{\fill}}lcccccccccccc}
\hline
\hline \cr
N$_{\rm cells}$ & $L_1$ & $R_1$ & $L_2$ & $R_2$ & $L_\infty$ & $R_\infty$ & $L_1$ & $R_1$ & $L_2$ & $R_2$ & $L_\infty$ & $R_\infty$ \cr \cline{2-13} \cr
& \multicolumn{6}{l}{\bf density-gas} & \multicolumn{6}{l}{\bf x-vel-gas} \cr
   16 &    4.2E-07 &    --  &    9.5E-07 &    --  &    2.7E-06 &    --  &    2.5E-06 &    --  &    5.7E-06 &    --  &    1.6E-05 &    --  \cr 
   32 &    1.1E-07 &    2.0 &    2.4E-07 &    2.0 &    6.8E-07 &    2.0 &    6.4E-07 &    2.0 &    1.4E-06 &    2.0 &    4.0E-06 &    2.0 \cr 
   64 &    2.7E-08 &    2.0 &    6.0E-08 &    2.0 &    1.7E-07 &    2.0 &    1.6E-07 &    2.0 &    3.6E-07 &    2.0 &    1.0E-06 &    2.0 \cr 
  128 &    6.8E-09 &    2.0 &    1.5E-08 &    2.0 &    4.3E-08 &    2.0 &    4.0E-08 &    2.0 &    9.0E-08 &    2.0 &    2.5E-07 &    2.0 \cr
\cline{2-7} \cline{8-13} \cr
& \multicolumn{6}{l}{\bf density-dust} & \multicolumn{6}{l}{\bf x-vel-dust} \cr
   16 &    2.5E-08 &    --  &    5.6E-08 &    --  &    1.6E-07 &    --  &    1.7E-06 &    --  &    3.7E-06 &    --  &    1.1E-05 &    --  \cr 
   32 &    4.8E-09 &    2.4 &    1.1E-08 &    2.4 &    3.0E-08 &    2.4 &    4.2E-07 &    2.0 &    9.3E-07 &    2.0 &    2.6E-06 &    2.0 \cr 
   64 &    3.6E-10 &    3.7 &    8.0E-10 &    3.7 &    2.4E-09 &    3.7 &    1.1E-07 &    2.0 &    2.3E-07 &    2.0 &    6.7E-07 &    2.0 \cr 
  128 &    3.5E-10 &    0.1 &    7.7E-10 &    0.1 &    2.2E-09 &    0.1 &    2.7E-08 &    2.0 &    5.9E-08 &    2.0 &    1.7E-07 &    2.0 \cr 
\hline
\hline
\end{tabular*}
\end{small}
\end{table}

\begin{table}  
\begin{small}
\caption{Convergence Rates: Case : $D=1.4\times 10^{-2},~{\bf k}=(1,0),~\rho_d/\rho_g=10^3$.\label{caseD.tab}}
\begin{tabular*}{\textwidth}{@{\extracolsep{\fill}}lcccccccccccc}
\hline
\hline \cr
N$_{\rm cells}$ & $L_1$ & $R_1$ & $L_2$ & $R_2$ & $L_\infty$ & $R_\infty$ & $L_1$ & $R_1$ & $L_2$ & $R_2$ & $L_\infty$ & $R_\infty$ \cr \cline{2-13} \cr
& \multicolumn{6}{l}{\bf density-gas} & \multicolumn{6}{l}{\bf x-vel-gas} \cr
   16 &    3.1E-07 &    --  &    6.9E-07 &    --  &    1.9E-06 &    --  &    5.5E-06 &    --  &    1.2E-05 &    --  &    3.4E-05 &    --  \cr 
   32 &    7.5E-08 &    2.0 &    1.7E-07 &    2.0 &    4.7E-07 &    2.0 &    1.4E-06 &    2.0 &    3.1E-06 &    2.0 &    8.7E-06 &    2.0 \cr 
   64 &    1.8E-08 &    2.1 &    4.0E-08 &    2.1 &    1.1E-07 &    2.1 &    3.5E-07 &    2.0 &    7.8E-07 &    2.0 &    2.2E-06 &    2.0 \cr 
  128 &    4.1E-09 &    2.1 &    9.2E-09 &    2.1 &    2.6E-08 &    2.1 &    8.8E-08 &    2.0 &    1.9E-07 &    2.0 &    5.5E-07 &    2.0 \cr
\cline{2-7} \cline{8-13} \cr
& \multicolumn{6}{l}{\bf density-dust} & \multicolumn{6}{l}{\bf x-vel-dust} \cr
   16 &    1.4E-06 &    --  &    3.0E-06 &    --  &    8.5E-06 &    --  &    5.5E-06 &    --  &    1.2E-05 &    --  &    3.4E-05 &    --  \cr 
   32 &    3.3E-07 &    2.0 &    7.4E-07 &    2.0 &    2.1E-06 &    2.0 &    1.4E-06 &    2.0 &    3.1E-06 &    2.0 &    8.7E-06 &    2.0 \cr 
   64 &    7.8E-08 &    2.1 &    1.7E-07 &    2.1 &    4.9E-07 &    2.1 &    3.5E-07 &    2.0 &    7.8E-07 &    2.0 &    2.2E-06 &    2.0 \cr 
  128 &    1.7E-08 &    2.2 &    3.8E-08 &    2.2 &    1.1E-07 &    2.2 &    8.8E-08 &    2.0 &    1.9E-07 &    2.0 &    5.5E-07 &    2.0 \cr
\hline
\hline
\end{tabular*}
\end{small}
\end{table}

\begin{table}  
\begin{small}
\caption{Convergence Rates: Case : $E=1.4\times 10^{-2},~{\bf k}=(1,0),~\rho_d/\rho_g=1$.\label{caseE.tab}}
\begin{tabular*}{\textwidth}{@{\extracolsep{\fill}}lcccccccccccc}
\hline
\hline \cr
N$_{\rm cells}$ & $L_1$ & $R_1$ & $L_2$ & $R_2$ & $L_\infty$ & $R_\infty$ & $L_1$ & $R_1$ & $L_2$ & $R_2$ & $L_\infty$ & $R_\infty$ \cr \cline{2-13} \cr
& \multicolumn{6}{l}{\bf density-gas} & \multicolumn{6}{l}{\bf x-vel-gas} \cr
   16 &    3.0E-06 &    --  &    6.8E-06 &    --  &    1.9E-05 &    --  &    2.6E-06 &    --  &    5.9E-06 &    --  &    1.7E-05 &    --  \cr 
   32 &    1.5E-06 &    1.0 &    3.3E-06 &    1.0 &    9.4E-06 &    1.0 &    1.1E-06 &    1.3 &    2.4E-06 &    1.3 &    6.7E-06 &    1.3 \cr 
   64 &    7.4E-07 &    1.0 &    1.6E-06 &    1.0 &    4.6E-06 &    1.0 &    4.0E-07 &    1.4 &    8.8E-07 &    1.4 &    2.5E-06 &    1.4 \cr 
  128 &    2.3E-07 &    1.7 &    5.1E-07 &    1.7 &    1.5E-06 &    1.7 &    1.2E-07 &    1.8 &    2.6E-07 &    1.8 &    7.4E-07 &    1.8 \cr
\cline{2-7} \cline{8-13} \cr
& \multicolumn{6}{l}{\bf density-dust} & \multicolumn{6}{l}{\bf x-vel-dust} \cr
   16 &    3.3E-05 &    --  &    7.3E-05 &    --  &    2.0E-04 &    --  &    2.7E-06 &    --  &    6.0E-06 &    --  &    1.7E-05 &    --  \cr 
   32 &    9.0E-06 &    1.9 &    2.0E-05 &    1.9 &    5.6E-05 &    1.9 &    1.0E-06 &    1.4 &    2.3E-06 &    1.4 &    6.5E-06 &    1.4 \cr 
   64 &    3.0E-06 &    1.6 &    6.7E-06 &    1.6 &    1.9E-05 &    1.6 &    3.6E-07 &    1.5 &    8.1E-07 &    1.5 &    2.3E-06 &    1.5 \cr 
  128 &    8.7E-07 &    1.8 &    1.9E-06 &    1.8 &    5.5E-06 &    1.8 &    1.0E-07 &    1.8 &    2.3E-07 &    1.8 &    6.6E-07 &    1.8 \cr 
\hline
\hline
\end{tabular*}
\end{small}
\end{table}


\begin{table}  
\begin{small}
\caption{Convergence Rates: Case : $F=1.4\times 10^{-2},~{\bf k}=(2,1)/\sqrt{5},~\rho_d/\rho_g=1 $.\label{caseF.tab}}
\begin{tabular*}{\textwidth}{@{\extracolsep{\fill}}lcccccccccccc}
\hline
\hline \cr
N$_{\rm cells}$ & $L_1$ & $R_1$ & $L_2$ & $R_2$ & $L_\infty$ & $R_\infty$ & $L_1$ & $R_1$ & $L_2$ & $R_2$ & $L_\infty$ & $R_\infty$ \cr \cline{2-13} \cr
& \multicolumn{6}{l}{\bf density-gas} & \multicolumn{6}{l}{\bf x-vel-gas} \cr
   16 &    2.1E-05 &    --  &    2.4E-05 &    --  &    3.3E-05 &    --  &    5.4E-05 &    --  &    6.0E-05 &    --  &    8.5E-05 &    --  \cr 
   32 &    6.0E-06 &    1.8 &    6.6E-06 &    1.8 &    9.4E-06 &    1.8 &    1.5E-05 &    1.9 &    1.6E-05 &    1.9 &    2.3E-05 &    1.9 \cr 
   64 &    1.6E-06 &    1.9 &    1.7E-06 &    1.9 &    2.5E-06 &    1.9 &    3.8E-06 &    2.0 &    4.2E-06 &    2.0 &    6.0E-06 &    2.0 \cr 
  128 &    4.0E-07 &    2.0 &    4.4E-07 &    2.0 &    6.3E-07 &    2.0 &    9.6E-07 &    2.0 &    1.1E-06 &    2.0 &    1.5E-06 &    2.0 \cr 
\cline{2-7} \cline{8-13} \cr
& \multicolumn{6}{l}{\bf y-vel-gas} & \multicolumn{6}{l}{\bf pressure} \cr
   16 &    1.7E-05 &    --  &    1.8E-05 &    --  &    2.6E-05 &    --  &    3.0E-05 &    --  &    3.3E-05 &    --  &    4.7E-05 &    --  \cr 
   32 &    4.8E-06 &    1.8 &    5.4E-06 &    1.8 &    7.6E-06 &    1.8 &    8.4E-06 &    1.8 &    9.3E-06 &    1.8 &    1.3E-05 &    1.8 \cr 
   64 &    1.3E-06 &    1.9 &    1.4E-06 &    1.9 &    2.0E-06 &    1.9 &    2.2E-06 &    1.9 &    2.4E-06 &    1.9 &    3.4E-06 &    1.9 \cr 
  128 &    3.3E-07 &    2.0 &    3.6E-07 &    2.0 &    5.1E-07 &    2.0 &    5.6E-07 &    2.0 &    6.2E-07 &    2.0 &    8.8E-07 &    2.0 \cr 
\cline{2-7} \cline{8-13} \cr
& \multicolumn{6}{l}{\bf density-dust} & \multicolumn{6}{l}{\bf x-vel-dust} \cr
   16 &    2.3E-04 &    --  &    2.5E-04 &    --  &    3.6E-04 &    --  &    7.8E-05 &    --  &    8.6E-05 &    --  &    1.2E-04 &    --  \cr 
   32 &    4.9E-05 &    2.2 &    5.5E-05 &    2.2 &    8.0E-05 &    2.2 &    2.0E-05 &    1.9 &    2.2E-05 &    1.9 &    3.2E-05 &    1.9 \cr 
   64 &    1.6E-05 &    1.7 &    1.7E-05 &    1.7 &    2.5E-05 &    1.7 &    5.1E-06 &    2.0 &    5.7E-06 &    2.0 &    8.0E-06 &    2.0 \cr 
  128 &    3.8E-06 &    2.1 &    4.3E-06 &    2.0 &    9.4E-06 &    1.4 &    1.3E-06 &    2.0 &    1.4E-06 &    2.0 &    2.0E-06 &    2.0 \cr 
\cline{2-7} \cline{8-13} \cr
& \multicolumn{6}{l}{\bf y-vel-dust} & \multicolumn{6}{l}{\bf  } \cr
   16 &    1.3E-04 &    --  &    1.5E-04 &    --  &    2.1E-04 &    --  &  &  &  &  &  &  \cr 
   32 &    3.5E-05 &    1.9 &    3.9E-05 &    1.9 &    5.5E-05 &    1.9 &  &  &  &  &  &  \cr 
   64 &    8.8E-06 &    2.0 &    9.8E-06 &    2.0 &    1.4E-05 &    2.0 &  &  &  &  &  &  \cr 
  128 &    2.2E-06 &    2.0 &    2.5E-06 &    2.0 &    3.5E-06 &    2.0 &  &  &  &  &  &  \cr 
\hline
\hline
\end{tabular*}
\end{small}
\end{table}

\begin{table}  
\begin{small}
\caption{Convergence Rates: Case : $G=1.4\times 10^{-2},~{\bf k}=(2,1)/\sqrt{5},~\rho_d/\rho_g=1 $.\label{caseG.tab}}
\begin{tabular*}{\textwidth}{@{\extracolsep{\fill}}lcccccccccccc}
\hline
\hline \cr
N$_{\rm cells}$ & $L_1$ & $R_1$ & $L_2$ & $R_2$ & $L_\infty$ & $R_\infty$ & $L_1$ & $R_1$ & $L_2$ & $R_2$ & $L_\infty$ & $R_\infty$ \cr \cline{2-13} \cr
& \multicolumn{6}{l}{\bf density-gas} & \multicolumn{6}{l}{\bf x-vel-gas} \cr
   16 &    5.8E-05 &    --  &    6.4E-05 &    --  &    9.7E-05 &    --  &    5.0E-05 &    --  &    5.6E-05 &    --  &    8.7E-05 &    --  \cr 
   32 &    1.3E-05 &    2.2 &    1.5E-05 &    2.1 &    2.4E-05 &    2.0 &    1.6E-05 &    1.6 &    1.8E-05 &    1.6 &    2.8E-05 &    1.6 \cr 
   64 &    3.7E-06 &    1.8 &    4.1E-06 &    1.9 &    6.3E-06 &    1.9 &    5.3E-06 &    1.6 &    6.0E-06 &    1.6 &    8.7E-06 &    1.7 \cr 
  128 &    1.5E-06 &    1.3 &    1.6E-06 &    1.3 &    2.6E-06 &    1.3 &    2.0E-06 &    1.4 &    2.3E-06 &    1.4 &    3.5E-06 &    1.3 \cr 
\cline{2-7} \cline{8-13} \cr
& \multicolumn{6}{l}{\bf y-vel-gas} & \multicolumn{6}{l}{\bf pressure} \cr
   16 &    2.5E-05 &    --  &    3.0E-05 &    --  &    7.0E-05 &    --  &    8.2E-05 &    --  &    9.0E-05 &    --  &    1.4E-04 &    --  \cr 
   32 &    3.2E-05 &   -0.4 &    3.6E-05 &   -0.2 &    5.6E-05 &    0.3 &    1.8E-05 &    2.2 &    2.1E-05 &    2.1 &    3.3E-05 &    2.0 \cr 
   64 &    2.5E-05 &    0.4 &    2.8E-05 &    0.4 &    3.9E-05 &    0.5 &    5.2E-06 &    1.8 &    5.8E-06 &    1.9 &    8.8E-06 &    1.9 \cr 
  128 &    1.5E-05 &    0.7 &    1.6E-05 &    0.7 &    2.3E-05 &    0.8 &    2.1E-06 &    1.3 &    2.3E-06 &    1.3 &    3.7E-06 &    1.3 \cr 
\cline{2-7} \cline{8-13} \cr
& \multicolumn{6}{l}{\bf density-dust} & \multicolumn{6}{l}{\bf x-vel-dust} \cr
   16 &    9.6E-05 &    --  &    1.1E-04 &    --  &    2.0E-04 &    --  &    4.9E-05 &    --  &    5.4E-05 &    --  &    8.1E-05 &    --  \cr 
   32 &    3.8E-05 &    1.3 &    4.4E-05 &    1.3 &    7.8E-05 &    1.3 &    1.8E-05 &    1.4 &    2.0E-05 &    1.4 &    2.9E-05 &    1.5 \cr 
   64 &    1.7E-05 &    1.1 &    1.9E-05 &    1.2 &    3.1E-05 &    1.3 &    6.1E-06 &    1.6 &    6.8E-06 &    1.6 &    9.8E-06 &    1.6 \cr 
  128 &    8.0E-06 &    1.1 &    9.1E-06 &    1.1 &    1.5E-05 &    1.1 &    2.2E-06 &    1.4 &    2.5E-06 &    1.4 &    3.8E-06 &    1.4 \cr 
\cline{2-7} \cline{8-13} \cr
& \multicolumn{6}{l}{\bf y-vel-dust} & \multicolumn{6}{l}{\bf  } \cr
   16 &    1.8E-04 &    --  &    2.0E-04 &    --  &    2.9E-04 &    --  &  &  &  &  &  &  \cr 
   32 &    8.1E-05 &    1.2 &    9.0E-05 &    1.2 &    1.3E-04 &    1.2 &  &  &  &  &  &  \cr 
   64 &    3.8E-05 &    1.1 &    4.2E-05 &    1.1 &    5.9E-05 &    1.1 &  &  &  &  &  &  \cr 
  128 &    1.8E-05 &    1.1 &    2.0E-05 &    1.1 &    2.8E-05 &    1.1 &  &  &  &  &  &  \cr 
\hline
\hline
\end{tabular*}
\end{small}
\end{table}

\begin{table}  
\begin{small}
\caption{Convergence Rates: Case : $H=1.4\times 10^{-2},~{\bf k}=(2,1)/\sqrt{5},~\rho_d/\rho_g=10^{-3} $.\label{caseH.tab}}
\begin{tabular*}{\textwidth}{@{\extracolsep{\fill}}lcccccccccccc}
\hline
\hline \cr
N$_{\rm cells}$ & $L_1$ & $R_1$ & $L_2$ & $R_2$ & $L_\infty$ & $R_\infty$ & $L_1$ & $R_1$ & $L_2$ & $R_2$ & $L_\infty$ & $R_\infty$ \cr \cline{2-13} \cr
& \multicolumn{6}{l}{\bf density-gas} & \multicolumn{6}{l}{\bf x-vel-gas} \cr
   16 &    1.5E-05 &    --  &    1.7E-05 &    --  &    2.4E-05 &    --  &    4.3E-05 &    --  &    4.8E-05 &    --  &    6.7E-05 &    --  \cr 
   32 &    4.5E-06 &    1.8 &    5.0E-06 &    1.8 &    7.1E-06 &    1.8 &    1.2E-05 &    1.9 &    1.3E-05 &    1.9 &    1.9E-05 &    1.8 \cr 
   64 &    1.2E-06 &    1.9 &    1.3E-06 &    1.9 &    1.9E-06 &    1.9 &    3.0E-06 &    2.0 &    3.4E-06 &    2.0 &    4.8E-06 &    2.0 \cr 
  128 &    3.0E-07 &    2.0 &    3.3E-07 &    2.0 &    4.8E-07 &    2.0 &    7.5E-07 &    2.0 &    8.4E-07 &    2.0 &    1.2E-06 &    2.0 \cr 
\cline{2-7} \cline{8-13} \cr
& \multicolumn{6}{l}{\bf y-vel-gas} & \multicolumn{6}{l}{\bf pressure} \cr
   16 &    1.1E-06 &    --  &    1.2E-06 &    --  &    1.8E-06 &    --  &    2.1E-05 &    --  &    2.4E-05 &    --  &    3.4E-05 &    --  \cr 
   32 &    5.4E-07 &    1.0 &    6.0E-07 &    1.1 &    8.6E-07 &    1.1 &    6.3E-06 &    1.8 &    7.0E-06 &    1.8 &    9.9E-06 &    1.8 \cr 
   64 &    2.0E-07 &    1.4 &    2.3E-07 &    1.4 &    3.3E-07 &    1.4 &    1.7E-06 &    1.9 &    1.8E-06 &    1.9 &    2.6E-06 &    1.9 \cr 
  128 &    6.4E-08 &    1.7 &    7.1E-08 &    1.7 &    1.0E-07 &    1.7 &    4.2E-07 &    2.0 &    4.7E-07 &    2.0 &    6.7E-07 &    2.0 \cr 
\cline{2-7} \cline{8-13} \cr
& \multicolumn{6}{l}{\bf density-dust} & \multicolumn{6}{l}{\bf x-vel-dust} \cr
   16 &    4.1E-07 &    --  &    4.6E-07 &    --  &    6.5E-07 &    --  &    1.8E-04 &    --  &    2.0E-04 &    --  &    2.8E-04 &    --  \cr 
   32 &    1.4E-07 &    1.6 &    1.5E-07 &    1.6 &    2.1E-07 &    1.6 &    4.9E-05 &    1.9 &    5.4E-05 &    1.9 &    7.6E-05 &    1.9 \cr 
   64 &    5.2E-08 &    1.4 &    5.8E-08 &    1.4 &    8.2E-08 &    1.4 &    1.2E-05 &    2.0 &    1.4E-05 &    2.0 &    1.9E-05 &    2.0 \cr 
  128 &    2.0E-08 &    1.4 &    2.2E-08 &    1.4 &    3.4E-08 &    1.3 &    3.1E-06 &    2.0 &    3.5E-06 &    2.0 &    4.9E-06 &    2.0 \cr 
\cline{2-7} \cline{8-13} \cr
& \multicolumn{6}{l}{\bf y-vel-dust} & \multicolumn{6}{l}{\bf  } \cr
   16 &    3.4E-05 &    --  &    3.7E-05 &    --  &    5.2E-05 &    --  &  &  &  &  &  &  \cr 
   32 &    9.9E-06 &    1.8 &    1.1E-05 &    1.8 &    1.6E-05 &    1.8 &  &  &  &  &  &  \cr 
   64 &    2.6E-06 &    1.9 &    2.9E-06 &    1.9 &    4.1E-06 &    1.9 &  &  &  &  &  &  \cr 
  128 &    6.7E-07 &    2.0 &    7.4E-07 &    2.0 &    1.0E-06 &    2.0 &  &  &  &  &  &  \cr 
\hline
\hline
\end{tabular*}
\end{small}
\end{table}

\begin{table}  
\begin{small}
\caption{Convergence Rates: Case : $I=1.4\times 10^{-2},~{\bf k}=(2,1)/\sqrt{5},~\rho_d/\rho_g=10^3 $.\label{caseI.tab}}
\begin{tabular*}{\textwidth}{@{\extracolsep{\fill}}lcccccccccccc}
\hline
\hline \cr
N$_{\rm cells}$ & $L_1$ & $R_1$ & $L_2$ & $R_2$ & $L_\infty$ & $R_\infty$ & $L_1$ & $R_1$ & $L_2$ & $R_2$ & $L_\infty$ & $R_\infty$ \cr \cline{2-13} \cr
& \multicolumn{6}{l}{\bf density-gas} & \multicolumn{6}{l}{\bf x-vel-gas} \cr
   16 &    6.5E-06 &    --  &    7.2E-06 &    --  &    1.0E-05 &    --  &    1.7E-04 &    --  &    1.9E-04 &    --  &    2.6E-04 &    --  \cr 
   32 &    2.3E-06 &    1.5 &    2.5E-06 &    1.5 &    3.5E-06 &    1.5 &    1.4E-04 &    0.3 &    1.6E-04 &    0.3 &    2.2E-04 &    0.2 \cr 
   64 &    1.0E-06 &    1.2 &    1.1E-06 &    1.2 &    1.6E-06 &    1.2 &    8.9E-05 &    0.7 &    9.9E-05 &    0.7 &    1.4E-04 &    0.7 \cr 
  128 &    5.3E-07 &    0.9 &    5.9E-07 &    0.9 &    8.3E-07 &    0.9 &    5.4E-05 &    0.7 &    6.0E-05 &    0.7 &    8.4E-05 &    0.7 \cr 
\cline{2-7} \cline{8-13} \cr
& \multicolumn{6}{l}{\bf y-vel-gas} & \multicolumn{6}{l}{\bf pressure} \cr
   16 &    2.3E-04 &    --  &    2.6E-04 &    --  &    3.6E-04 &    --  &    9.1E-06 &    --  &    1.0E-05 &    --  &    1.4E-05 &    --  \cr 
   32 &    2.0E-04 &    0.2 &    2.2E-04 &    0.2 &    3.1E-04 &    0.2 &    3.2E-06 &    1.5 &    3.5E-06 &    1.5 &    5.0E-06 &    1.5 \cr 
   64 &    1.2E-04 &    0.7 &    1.4E-04 &    0.7 &    2.0E-04 &    0.6 &    1.4E-06 &    1.2 &    1.5E-06 &    1.2 &    2.2E-06 &    1.2 \cr 
  128 &    7.5E-05 &    0.7 &    8.3E-05 &    0.7 &    1.2E-04 &    0.7 &    7.4E-07 &    0.9 &    8.2E-07 &    0.9 &    1.2E-06 &    0.9 \cr 
\cline{2-7} \cline{8-13} \cr
& \multicolumn{6}{l}{\bf density-dust} & \multicolumn{6}{l}{\bf x-vel-dust} \cr
   16 &    2.2E-05 &    --  &    2.4E-05 &    --  &    3.3E-05 &    --  &    3.8E-04 &    --  &    4.1E-04 &    --  &    5.8E-04 &    --  \cr 
   32 &    9.1E-06 &    1.2 &    1.0E-05 &    1.2 &    1.4E-05 &    1.2 &    2.0E-04 &    0.9 &    2.3E-04 &    0.9 &    3.2E-04 &    0.9 \cr 
   64 &    4.0E-06 &    1.2 &    4.4E-06 &    1.2 &    6.2E-06 &    1.2 &    1.1E-04 &    0.9 &    1.2E-04 &    0.9 &    1.7E-04 &    0.9 \cr 
  128 &    1.9E-06 &    1.1 &    2.1E-06 &    1.1 &    3.0E-06 &    1.1 &    5.8E-05 &    0.9 &    6.5E-05 &    0.9 &    9.2E-05 &    0.9 \cr 
\cline{2-7} \cline{8-13} \cr
& \multicolumn{6}{l}{\bf y-vel-dust} & \multicolumn{6}{l}{\bf  } \cr
   16 &    5.2E-04 &    --  &    5.8E-04 &    --  &    8.1E-04 &    --  &  &  &  &  &  &  \cr 
   32 &    2.9E-04 &    0.9 &    3.2E-04 &    0.9 &    4.5E-04 &    0.8 &  &  &  &  &  &  \cr 
   64 &    1.5E-04 &    0.9 &    1.7E-04 &    0.9 &    2.4E-04 &    0.9 &  &  &  &  &  &  \cr 
  128 &    8.2E-05 &    0.9 &    9.1E-05 &    0.9 &    1.3E-04 &    0.9 &  &  &  &  &  &  \cr 
\hline
\hline
\end{tabular*}
\end{small}
\end{table}

\begin{table}  
\begin{small}
\caption{Convergence Rates: Case : $L=1.4\times 10^{-2},~{\bf k}=(2,1)/\sqrt{5},~\rho_d/\rho_g=1 $.\label{caseL.tab}}
\begin{tabular*}{\textwidth}{@{\extracolsep{\fill}}lcccccccccccc}
\hline
\hline \cr
N$_{\rm cells}$ & $L_1$ & $R_1$ & $L_2$ & $R_2$ & $L_\infty$ & $R_\infty$ & $L_1$ & $R_1$ & $L_2$ & $R_2$ & $L_\infty$ & $R_\infty$ \cr \cline{2-13} \cr
& \multicolumn{6}{l}{\bf density-gas} & \multicolumn{6}{l}{\bf x-vel-gas} \cr
   16 &    3.2E-05 &    --  &    3.5E-05 &    --  &    4.9E-05 &    --  &    4.3E-05 &    --  &    4.8E-05 &    --  &    6.7E-05 &    --  \cr 
   32 &    8.1E-06 &    2.0 &    9.0E-06 &    2.0 &    1.3E-05 &    1.9 &    4.4E-05 &   -0.0 &    4.9E-05 &   -0.0 &    6.9E-05 &   -0.0 \cr 
   64 &    1.6E-05 &   -1.0 &    1.8E-05 &   -1.0 &    2.5E-05 &   -1.0 &    4.3E-05 &    0.0 &    4.8E-05 &    0.0 &    6.8E-05 &    0.0 \cr 
  128 &    8.8E-06 &    0.9 &    9.7E-06 &    0.9 &    1.4E-05 &    0.9 &    1.9E-05 &    1.2 &    2.1E-05 &    1.2 &    3.0E-05 &    1.2 \cr 
\cline{2-7} \cline{8-13} \cr
& \multicolumn{6}{l}{\bf y-vel-gas} & \multicolumn{6}{l}{\bf pressure} \cr
   16 &    2.6E-05 &    --  &    2.9E-05 &    --  &    4.1E-05 &    --  &    4.4E-05 &    --  &    4.9E-05 &    --  &    6.9E-05 &    --  \cr 
   32 &    8.1E-05 &   -1.6 &    9.0E-05 &   -1.6 &    1.3E-04 &   -1.6 &    1.1E-05 &    2.0 &    1.3E-05 &    2.0 &    1.8E-05 &    2.0 \cr 
   64 &    5.9E-05 &    0.5 &    6.6E-05 &    0.5 &    9.3E-05 &    0.5 &    2.2E-05 &   -1.0 &    2.5E-05 &   -1.0 &    3.5E-05 &   -1.0 \cr 
  128 &    2.4E-05 &    1.3 &    2.7E-05 &    1.3 &    3.8E-05 &    1.3 &    1.2E-05 &    0.9 &    1.4E-05 &    0.9 &    1.9E-05 &    0.9 \cr 
\cline{2-7} \cline{8-13} \cr
& \multicolumn{6}{l}{\bf density-dust} & \multicolumn{6}{l}{\bf x-vel-dust} \cr
   16 &    1.3E-04 &    --  &    1.4E-04 &    --  &    2.0E-04 &    --  &    5.2E-05 &    --  &    5.7E-05 &    --  &    8.0E-05 &    --  \cr 
   32 &    1.6E-04 &   -0.3 &    1.7E-04 &   -0.3 &    2.5E-04 &   -0.3 &    4.1E-05 &    0.3 &    4.6E-05 &    0.3 &    6.5E-05 &    0.3 \cr 
   64 &    1.1E-04 &    0.6 &    1.2E-04 &    0.5 &    1.7E-04 &    0.5 &    4.4E-05 &   -0.1 &    4.8E-05 &   -0.1 &    6.9E-05 &   -0.1 \cr 
  128 &    4.4E-05 &    1.3 &    4.9E-05 &    1.3 &    7.0E-05 &    1.3 &    2.0E-05 &    1.2 &    2.2E-05 &    1.2 &    3.1E-05 &    1.2 \cr 
\cline{2-7} \cline{8-13} \cr
& \multicolumn{6}{l}{\bf y-vel-dust} & \multicolumn{6}{l}{\bf  } \cr
   16 &    1.8E-04 &    --  &    2.0E-04 &    --  &    2.8E-04 &    --  &  &  &  &  &  &  \cr 
   32 &    1.2E-04 &    0.5 &    1.4E-04 &    0.5 &    1.9E-04 &    0.5 &  &  &  &  &  &  \cr 
   64 &    6.9E-05 &    0.8 &    7.7E-05 &    0.8 &    1.1E-04 &    0.8 &  &  &  &  &  &  \cr 
  128 &    2.7E-05 &    1.4 &    3.0E-05 &    1.4 &    4.2E-05 &    1.4 &  &  &  &  &  &  \cr 
\hline
\hline
\end{tabular*}
\end{small}
\end{table}

\subsection{Streaming Instability} \label{strins:se}

\begin{figure} 
\begin{center}
\includegraphics[width=1\textwidth,scale=1.0]{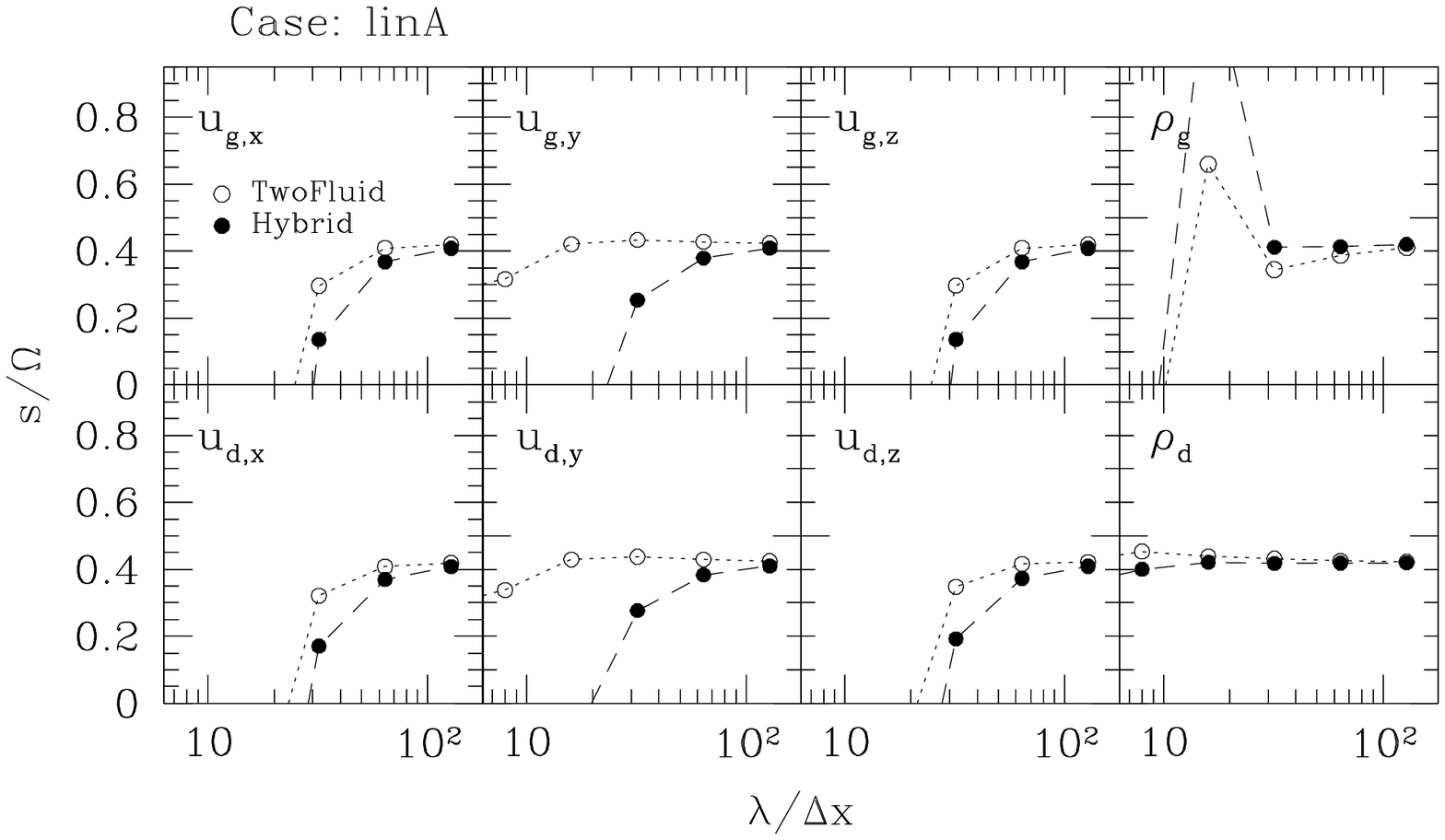}
\includegraphics[width=1\textwidth,scale=1.0]{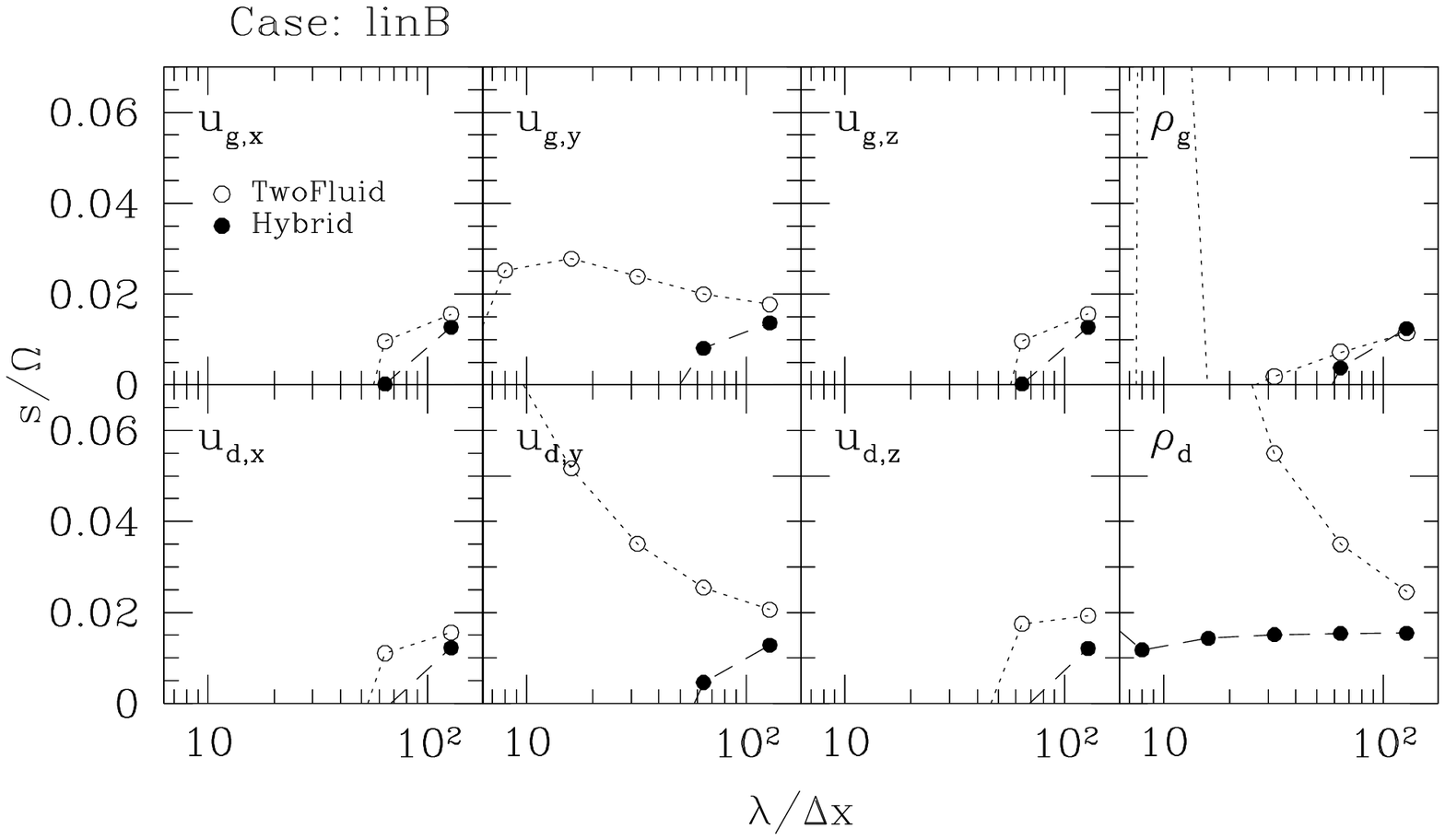}
\caption{ 
\label{si:fig}}
\end{center}
\end{figure}
In this last section we carry out the test proposed in~\cite{yojo07}
(hereafter, YJ07), which consists of following the growth of an
initial perturbation unstable to the streaming
instability~\cite{yogo05}.  The test setup, while considerably
simplified with respect to a Keplerian disk, preserves the key
dynamical features leading to the streaming instability in a
protoplanetary disk~\cite{johansen06,johansen07} and, with the
available analytic solution (for the linear regime), provides a clean
test for a simulation code.  The test consists of simulating an
axisymmetric shearing sheet in which dust particles drift with respect
to the gas due to the radial pressure gradient affecting the gas but
not the dust.  Both vertical structure (along the y-direction) and
self-gravity are ignored.  After neglecting the vertical derivatives
(along the y-axis), the governing equations for the gas component in a
shearing sheet can be written as in Eq.~(\ref{hypsys:eq}) provided
that we replace $f_d$ in the source term in~(\ref{fluid:eq}) with
\begin{equation}
f_d^\Omega = f_d + \Omega
\begin{pmatrix}
2(u_{g,y}+\eta v_k)\\
-\frac{1}{2}u_{g,x}\\
0
\end{pmatrix},
\end{equation}
where $\Omega$ is the angular velocity of the shearing sheet, $v_k=\Omega r$
is the azimuthal velocity at radial distance $r$, and $\eta$ is a 
dimensionless parameter expressing the strength of the radial pressure 
gradient with respect to the centrifugal force. 
The source term for the dust fluid is modified in an analogous way,
except that in this case we set $\eta=0$.
Finally, the particle equations of motion are also modified, 
namely Eq.~(\ref{dvdt:eq}) now reads
\begin{equation}
\frac{dv_d}{dt} = \omega\cdot v_d- \kd \,(v_{d}-u_{g}),\quad
\omega=
\begin{pmatrix}
0 & 2&0\\
-\frac{1}{2}&0&0\\
0 &0&0
\end{pmatrix}.
\end{equation}
In principle the y-component of the particles position should also be
modified, but this is irrelevant due to the azimuthal symmetry of the
system.

In order to perform the test, the gas and dust velocities are
initialized to the equilibrium solution given by Nakagawa et
al.~\cite{naseha86} (or Eq. 7 in YJ05).
Gas density and pressure are set according to
\begin{gather}
\rho_g = 1.4,\quad P=\frac{\rho_g}{\gamma x_p},
\end{gather}
where $x_p\equiv\eta v_k/c= 0.05$ as in YJ05, and $\rho_d$ is set
according to a specified dust-to-gas ratio (see below).  Note that the
parameters $\Omega,~r$ and $\eta$ need not be specified as long as one
expresses time, distances and velocities in units of
$\Omega^{-1},~\eta r$, and $\eta\Omega r=\eta v_k$, respectively.
Finally, a perturbation unstable to streaming instability, is added to
the initial equilibrium values.  YJ07 provide two sets of the
eigenmodes for the perturbation amplitudes of each variable,
corresponding to what they refer to as linA and linB cases.  Some
basic features of these cases, including the wave vector, the
dust-to-gas ratio and the normalized growth rate ($s/\Omega$), are
summarized in Table~\ref{siset:tab}, full details are provided in
YJ07. The perturbations are in the radial and vertical directions
(x,z) but not in the azimuthal direction (y). The system is evolved
using an isothermal equation of state. Note that these tests are
basically in the non-stiff regime, as $k_{g,d}\Delta t\le 0.1$.

In Fig.~\ref{si:fig} we report the simulated growth rate of the
instability, for both gas and dust variables, as a function of the
resolution expressed in number of cells per perturbation
wavelength. The top and bottom panels correspond to case linA and
linB, respectively, and open and filled dots refer to the two-fluid
and hybrid algorithm, respectively.  With respect to the hybrid
algorithm, we always employ 16 particles per cell and a TSC
interpolation scheme.
As for the linA case, the test
indicates that for both the two-fluid and hybrid algorithms, at least
32 cells per wavelength are required in order to capture the
instability growth for all variables and convergence to the analytic
solution is basically achieved with 64 cells per wavelength.  The linB
case is much more challenging and now at least 128 cells per
wavelength are necessary in order to capture the instability.  
 Note so that because the gas-dust coupling is non-stiff,
we do not
expect any particular advantage of our scheme with respect to other
second order schemes. In fact, our results are comparable to those
of Balsara et al.~\cite{balsara09}, who use an approach based on
a second order Godunov's method, and less accurate then those
in YJ05, who instead use a sixth order spectral method.

\begin{table}  
\caption{Run Set\label{siset:tab}}
\begin{tabular*}{\textwidth}{@{\extracolsep{\fill}}lcccc}
\hline \hline 
case & $\kappa_0$ & ${\bf k}$ & $\rho_d/\rho_g$ & $s/\Omega$  \cr
\hline
linA& $0.357$ & $(30,0,30)$ & $ 3$ & 0.4190204 \cr
linB& $0.357$ & $(6,0,6)$  & $0.2$ & 0.0154764 \cr  
\hline
\end{tabular*}
\end{table}
\section{Discussion \& Summary}\label{concl:se}
We have presented a stable and convergent method for studying a
system of gas and dust coupled through viscous drag in both non-stiff
and stiff regimes. 
Our approach consists of updating the fluid quantities using a two
fluid model and then using the updated fluid solution to advance the
individual particle solutions with a self-consistent time
evolution of the gas velocity in the estimate of the drag force.

In our derivation of the two fluid method, we first obtain a fluid
description of the dust component using a Particle-Mesh method, and
then we study the modified gas-dust hyperbolic system following the
approach in Miniati \& Colella~\cite{mico07a}.  Based on this analysis
we formulate a predictor step providing first order accurate
reconstruction of the time-averaged state variables at cell
interfaces, whence a second order accurate estimates of the
conservative fluxes can be obtained.  Finally, for the
time-discretization for the source terms we use a single-step,
second-order accurate scheme derived from the $\alpha-$QSS method
proposed by Mott, Oran and van Leer (\cite{Mottetal2000}). This
completes the description of our two fluid method.

The fluid description of the dust component (Eq. \ref{drddt:eq},~\ref{dvddt:eq})
assumes the simplest type 
of closure which neglects dispersion velocity terms in the momentum equation.
However, since the particle distribution is known, different and more suitable 
closures can be constructed according to the specifics of the aimed application.
This is relevant when the dust-gas coupling is stiff and the 
dust backreaction is important, e.g. high dust density and large dust grain
size~\cite{garaudetal2004}, because in this case the gas tends to follow the dust and
the anisotropic particle motions may not be quickly damped.

In order to advance the individual particle solutions we use the fluid
solution to determine the time dependence of the gas velocity entering
the drag terms.  This allows us to derive a particle
integration scheme, also based on the $\alpha$-QSS method, which is
second order accurate regardless of the strength of dust-gas coupling.
Remarkably, the particle method that we employ contains no explicit
term arising from the stiff coupling of the particle component, in the
sense that each particle motion is integrated individually, though it
effectively depends on the other particles solutions, and the scheme
is essentially explicit in time, as it only involves the particle
solution at time $t=n\Delta t$.
Note that when the dust backreaction on the gas is 
not important (low dust density), the particle scheme presented in Sec.~\ref{pis:se} 
can be used together with an unmodified Godunov's method for the gas
component. Similarly, a simpler scheme can be obtained 
in the limit $\kg\rightarrow 0$, as the gas description of
the semi-implicit two-fluid scheme in Sec.~\ref{sdc:se}
reduces to an ordinary explicit Godunov's method, while the stiff solver
remains unchanged for the dust component.

A set of benchmark problems show that our method is stable and
convergent. In particular, the two-fluid approach is second order
accurate both in the non-stiff and stiff regimes, and drops to first
order in the intermediate regime as expected
theoretically~\cite{mico07a}.  The hybrid scheme, on the other hand,
is second order only in the non-stiff regime. Since, as illustrated in
the Sec.~\ref{convrate_ps:se} and~\ref{tfconvrate:se}, both the
particle scheme and the two-fluid scheme retain their second order
accuracy irrespective of the stiffness conditions, the drop in
accuracy in the hybrid approach is most likely due to the difficulty of
coupling the gas and the dust fully self-consistently in the stiff
regimes.  At any rate, the scheme remains stable and first order
convergent even in the stiff regime.

We have also tested our code against the streaming instability
problem presented in YJ05. This is a clean test and a very relevant one
for protoplanetary disk applications. However, due to the specifics of
the test set up, the gas-dust coupling is non-stiff, so that we do not
expect any particular advantage of our scheme with respect to other
second order schemes. In fact, our results are comparable to those
of Balsara et al.~\cite{balsara09}, who use an approach based on
a second order Godunov's method, and less accurate then those
in YJ05, who instead use a sixth order spectral method.

Finally, although the present analysis focuses on the Epstein regime for 
the functional form of drag force, in principle extension to the case of Stokes 
regime should also be possible. This would require a modification of both the
Godunov predictor step and the semi-implicit scheme in Sec.~\ref{sim:se}.
In addition, the present scheme can also be extended to the case of dust 
particles of multiple sizes.  A complication arises, however, due to the 
fact that one cannot straightforwardly define a single effective dust 
component to which the gas is coupled via the drag force.  Such complication 
seems inevitable and of general character, i.e. it is not restricted 
to the approach presented in this paper, when the coupling is stiff
and the dust backreaction is dynamically important. So, for a limited
number of dust components of different grain size the present method
can be modified by defining an extended system in which each dust 
component is represented separately. This approach, however, becomes 
cumbersome and expensive when the number of dust species is large, and
a method for constructing a single effective dust component should be 
investigated in such cases.

\vskip 1truecm
\leftline{\bf Acknowledgment}
I acknowledge very useful and encouraging discussions with P. Colella about
various algorithmic issues, with L. Fouchet, A. Booley about
protoplanetary disks and A. Youdin, A. Johanson and H. Klahr about the
streaming instability test. I am thankful to an anonymous referee for useful comments
to the manuscript. This work was supported by ETH through a
Zwicky Prize Fellowship program.

\appendix
\section{Derivation of the $\alpha$-QSS Based Particle Scheme} \label{paqss:app}

We use the alpha-QSS method to integrate the following equations,
describing the trajectories in phase-space of dust particles,
\begin{eqnarray} \label{dxdt:aeq} 
\frac{d x_d}{dt} &=& v_d, \\ \label{dvdt:aeq}
\frac{dv_d}{dt}&=& - \kd \,(v_{d}-u_{g}) -\nabla \phi,
\end{eqnarray}
from time $t=t^n$, when $x_d(t)=x^n,~v_d(t)=v_d^n$, to $t=t^{n+1}=t^n+\Delta t$. 
Following Eq. (\ref{qsspred:eq}) we first integrate Eq. (\ref{dvdt:aeq}) assuming
$p_0=\kd^n$ and $q_0=\kd^nu_g^n-\nabla \phi$, which yields
\begin{equation}
\tilde v_d(t)=v_d^n + (u_g^n-v_d^n)(1-e^{-\kd^n t}) -t\, \nabla\phi.
\end{equation}
Note that we have modified the gravitational acceleration term because, 
as discussed in Sec.~\ref{sim:se}, gravitational acceleration is not attenuated 
by drag as both gas and dust are equally accelerated by it.
We then obtain a first order accurate expression of the particle position by time
integration of the above solution,
\begin{eqnarray}
\tilde x_d(\Delta t) &=& x_d^n+\int_t^{t+\Delta t} d\tau \tilde v_d(\tau)= \nonumber\\
&= & x_d^n+ \Delta t v_d^n+ \Delta t(u_g^n-v_d^n)\left(1-\frac{1-e^{-\kd \Delta t}}{\kd \Delta t}\right)
-\frac{\Delta t^2}{2}\, \nabla\phi,
\end{eqnarray}
which is equivalent to Eq. (\ref{xtlde:eq}).
For the final corrector step we need to estimate the equivalent of $q_*$ and $\bar p$
appearing in (\ref{qsscorr:eq}). The expressions for 
$q_*$ and and $\bar p$ in Eq. (\ref{qssqparams:eq}) and (\ref{qsspparams:eq}), respectively,
can be obtained by assuming a linear time dependence for these parameters. Instead for the
gas velocity (which play the equivalent role of $q$ in our case) we
use the {\it ansatz} in Eq. (\ref{vgoft:eq}) with 
$$\Delta v_g\equiv v_g^{n+1}- v_g^{n} .$$
On the other hand, for the drag coefficients (which play the equivalent role of $p$ in our case)
we use the following averages consistent with the prescription in Eq.  (\ref{qsspparams:eq}),
$$\kappa_s=\frac{1}{2}[\kappa_s(x_d^n)+\kappa_s(\tilde x_d)],~s=d,g.$$
Using Eq. (\ref{vgoft:eq}) and the above expressions for $\Delta v_g$ and $\kd,~\kg$,
we can then integrate Eq. (\ref{dvdt:aeq}) to obtain a time dependent solution of 
the dust particle velocity. Its evaluation at time $t^{n+1}$ gives the solution in 
Eq. (\ref{DvDt:eq}) and its integration over a timestep $\Delta t$ gives the solution
(\ref{DxDt:eq}).

\bibliographystyle{plain}
\bibliography{books,codes,papers,proceed}

\end{document}